\documentclass[prd,eqsecnum,nofootinbib,twocolumn,noeprint]{revtex4-2}
\usepackage{bm,amssymb,amsmath,amsfonts}
\usepackage{xcolor}
\usepackage{comment}
\usepackage{multirow}
\usepackage{pagecolor}
\usepackage{comment}
\usepackage{tikz}
\usepackage{mathrsfs}
\usepackage{orcidlink}
\usepackage[scr=esstix]{mathalfa} 
\usetikzlibrary{arrows.meta}

\usepackage{hyperref}
\hypersetup{
    colorlinks=true,
    linktocpage,
    unicode,
    linkcolor=red,
    filecolor=magenta,      
    urlcolor=cyan,
    citecolor=blue
}
\definecolor{red}{rgb}{1,0,0}
\definecolor{blue}{rgb}{0,0,1}
\definecolor{green}{rgb}{0,1,0}
\usetikzlibrary{calc}

\newcommand{\ud}{\mathrm{d}}    

\newcommand{\Lie}{\mathscr{L}} 
\newcommand{\gF}{\mathscr{F}} 
\newcommand{\Jtf}{\mathscr{J}} 

\renewcommand{\Re}{\operatorname{Re}}
\renewcommand{\Im}{\operatorname{Im}}

\begin{document}

\title{Octupole moments and the non-universality of free-fall in general relativity}

\author{Abraham I. Harte,\,\orcidlink{0000-0002-5893-5680}}
\, \email{abraham.harte@dcu.ie}

\affiliation{Centre for Astrophysics and Relativity (CfAR), School of Mathematical Sciences, \\ Dublin City University, Glasnevin, Dublin 9, Ireland}

\author{Paul Ramond\,\orcidlink{0000-0001-7123-0039}}\,\email{ramond@lpccaen.in2p3.fr}

\affiliation{Laboratoire de Physique Corpusculaire (LPC) Caen, Université de Caen Normandie,\\ ENSICAEN, CNRS/IN2P3, LPC Caen UMR6534, F-14000 Caen, France}

\date{\today}

\begin{abstract}
    Extended bodies in general relativity do not necessarily fall along geodesics. Instead, a body can experience accelerations that depend on its angular momentum, as well as on its quadrupole, octupole, and higher-order moments. Moreover, due to the fact that one object's multipole moments can evolve differently from those of another, different bodies can fall differently---even with identical initial data. This paper examines how octupole moments contribute to the non-universality of free-fall in general relativity. We begin by showing that in arbitrary vacuum spacetimes, only the trace-free component of an octupole moment can affect an object's motion. It follows that at least 16 out of 40 octupole components decouple from the laws of motion. We also obtain two decompositions for trace-free octupole moments, one in terms of a timelike frame vector and the other in terms of a null tetrad. These decompositions are applied to motion both in static  Newtonian spacetimes and in fully-relativistic vacuum spacetimes that are of Petrov type D (such as Kerr). In Newtonian spacetimes, the mass moments are shown to have their ordinary Newtonian effects. The momentum moments instead determine a body's hidden momentum---the misalignment between its momentum and its velocity. In Petrov type D spacetimes, we show that some torques that are impossible with quadrupole moments are generically possible with octupole moments. Octupole moments can thus have qualitatively-different effects from quadrupole moments. 
\end{abstract}

\maketitle

\section{Introduction}

To a first approximation, freely-falling bodies in general relativity move on geodesics while their spins are parallel transported. However, corrections can arise for two reasons: self-interaction and finite size. All objects perturb the spacetime geometry, and an object's interaction with its own perturbations can affect its motion \cite{Po.al.11, Ha.15, BaPo.18}. This ``self-interaction'' can produce both a ``self-force'' and a ``self-torque,'' and can also ``dress'' an object's mass and other characteristics with the inertia of an object's self-field. The self-force is, e.g., what drives the inspiral of compact binaries. No analogous effects occur in Newtonian gravity, where the self-force, the self-torque, and all dressings identically vanish. 

Gravitational self-interaction depends on a body's long-ranged gravitational field. Extended-body effects instead depend on the short-ranged matter fields associated with a body's internal structure \cite{Di.79, ThHa.85}. Extended-body effects arise even in Newtonian gravity, where they can drive tidal locking in binaries and produce secular changes in orbital parameters \cite{Hu.81, Tr.23, PoWi, Og.14}. 
If self-interaction is ignored and an object is sufficiently small compared with all external lengthscales, its structure does not affect its motion: All objects with the same initial positions and the same initial velocities fall on the same trajectories. In Newtonian gravity, the first corrections to this result arise at quadrupolar order, involving force differences that are suppressed by a factor of order $[\mbox{(body size)/(external scale)}]^2$. The relativistic theory additionally admits dipole corrections associated with an object's angular momentum. However, those corrections are still ``universal'' in the sense that all objects with the same initial position, the same initial velocity, the same initial angular momentum fall identically when all higher moments are neglected. In both the Newtonian and the relativistic theories, the universality of free-fall is broken at quadrupolar order. This is because the quadrupole and higher-order moments can evolve differently depending on, e.g., a body's composition. Understanding such differences is the primary motivation for this paper.

Building upon previous work at (leading) quadrupolar order \cite{Ha.20, Ha.23}, we discuss (subleading) octupolar forces and torques in general relativity, ignoring self-interaction. One motivation for considering octupole moments is that understanding the effects of additional moments can allow more of an object's internal structure---and perhaps its composition---to be inferred by observing its motion. It has, for example, been suggested that the non-universality of free-fall can be used to constrain the equations of state relevant for neutron stars \cite{FlHi.08}. In that context, knowledge of (say) an octupole moment could provide additional information that places more stringent constraints on an object's interior. It is, however, essential to understand what exactly can be determined by observing an object's motion: Which components of a particular multipole moment couple to the motion, and of those components, which have degenerate contributions?

Another motivation for considering octupole moments is to determine if they can produce effects that are qualitatively different from any possible quadrupolar effects. In order to understand this most clearly, it can be convenient not to imagine the motion of a naturally-occurring astrophysical system, but rather that of a spacecraft that has been designed to control its shape---and thus its multipole moments. Such a spacecraft can generically maneuver without the use of a rocket \cite{Wi.03, Ha.07, Ha.20, Silva.al.16, Silva.22}. While some maneuvers are possible even in Newtonian gravity \cite{Ma.87, La.92, Be.01, GrTu.03, Ha.21}, additional effects are sometimes possible in relativistic settings. In both cases, however, there are restrictions; some motions are not possible for any sequence of shapes. It has long been known, for example, that at least some constraints are imposed by Killing fields \cite{DiI.70}. The existence of rotational Killing fields would, e.g., preclude an object from changing its orbital angular momentum without a compensating change in its spin angular momentum.

Not all constraints arise from Killing fields, however. Additional restrictions are known to appear at quadrupolar order \cite{Ha.20}, where they can be related to a weak type of ``local symmetry'' of the Weyl tensor that generically exists at least in algebraically-special spacetimes \cite{Ha.23}. As a special case, it is impossible for the quadrupole moment of any extended body in the Kerr spacetime to produce a torque $N^{ab} = N^{[ab]}$ in which $N^{ab} f_{ab} \neq 0$, where $f_{ab}$ denotes any conformal Killing-Yano tensor. This reduces the space of possible torques from six to four. More generally, there are various constraints on quadrupolar forces and torques that depend only on the Petrov type of the background Weyl tensor. One question addressed here is whether or not such constraints persist at higher orders. At least in the cases we consider, they do not. Independent of any material model, we provide examples in which octupole moments can have qualitatively different effects from quadrupole moments.

As a technical issue, octupole moments---which are rank-5 tensors with nontrivial index symmetries---can be difficult to work with directly. To alleviate this, we provide two simplifying decompositions. Most familiarly, we find a $3+1$ decomposition for the trace-free octupole moment in terms of a trace-free mass octupole and a trace-free momentum octupole. Building upon work in the quadrupole case, we also provide a decomposition with respect to a null tetrad, perhaps one that has been aligned with a spacetime's principal null directions. This allows any trace-free octupole moment to be written entirely in terms of three complex vectors.

We begin in Sec. \ref{Sec:review} by reviewing relevant aspects of Dixon's formalism \cite{Di.74, Di.79, Di.15, Ha.15} for extended-body motion in general relativity. Sec. \ref{Sec:NoTrace} then shows that traces of a body's octupole moment cannot affect its motion in any spacetime that satisfies the vacuum Einstein equation. This is similar to what is known to occur in the quadrupole case \cite{Ha.10, BiGe.14, Ha.20}. We show, however, that the pattern does not extend to the hexadecapole moment, whose traces \textit{can} affect a body's motion. Next, Sec. \ref{Sec:31decomp} performs $3+1$ decompositions of both the full and the trace-free octupole moments. Sec. \ref{Sec:NullDecomp} instead decomposes the trace-free octupole in terms of a null tetrad. Sec. \ref{Sec:Newtonian} applies our $3+1$ decomposition to understand forces and torques that can act on (not necessarily Newtonian) extended bodies in nearly-Newtonian spacetimes. Lastly, our null decomposition is applied in Sec. \ref{Sec:TypeD} to describe forces and torques in fully-relativistic type D spacetimes. In both contexts, we find that some effects forbidden at quadrupolar order can occur at octupolar order.

There are three appendices. App. \ref{App:Notation} summarizes our notation and conventions. It also  includes a table of symbols. App. \ref{App:TF} obtains trace-free projections for rank-5 tensors with the properties of an octupole moment. Finally, App. \ref{App:inducedMoments} applies our decompositions to spin-induced multipole moments that have appeared in the literature.

\section{Review of extended-body motion in curved spacetimes}
\label{Sec:review}

Our goal in this paper is to understand the motion of uncharged extended bodies in general relativity. We assume that any such body has finite size, and is therefore  described by a spatially-compact stress-energy tensor $T^{ab} = T^{(ab)}$. If there are no external (non-gravitational) forces, and if matter is neither lost nor accreted, an object's stress-energy tensor must be conserved:
\begin{equation}
    \nabla_b T^{ab} = 0.
    \label{stressCons}
\end{equation} 
We do not attempt to model the full dynamics of $T^{ab}$, but instead focus on a body's bulk characteristics: its linear momentum, its angular momentum, and perhaps a suitable centroid. Useful definitions for these characteristics have been identified and analyzed by Dixon \cite{Di.74, Di.79, Di.15}, and this section reviews relevant aspects of his formalism. 

\subsection{Overview of Dixon's formalism}

\begin{figure}

\includegraphics[width=.45\linewidth]{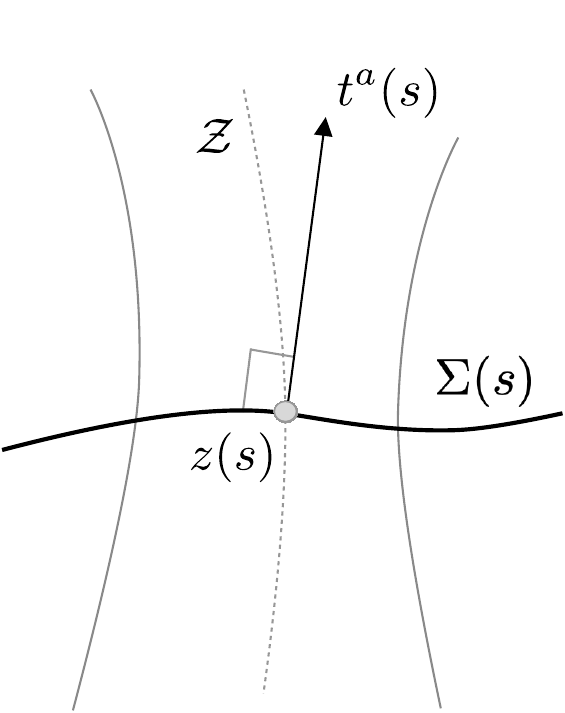}

\caption{The worldtube of an extended body and the geometry of the multipolar decomposition. In order to describe an extended body using Dixon's multipole moments, one must fix both a reference worldline $\mathcal{Z}$ and a foliation of the worldtube. Parameterizing that worldline by $z(s)$, each leaf $\Sigma(s)$ of the foliation is to be formed from the union of all geodesics that pass through $z(s)$ and are orthogonal at that point to a given timelike vector $t^a(s)$.}

\label{Fig:worldtube}
\end{figure}

Dixon's work on extended bodies may be viewed as a theory of multipole moments for conserved, spatially-compact stress-energy tensors in curved spacetimes\footnote{Although we make no use of it here, Dixon also allowed for electromagnetic interactions \cite{Di.74}. In that context, $\nabla_b T^{ab} = F^{ab} J_b$, where $F_{ab} = F_{[ab]}$ denotes an electromagnetic field and $J_a$ an object's current density. So-called ``reduced'' multipole moments can then be developed for both $T^{ab}$ and for $J_a$.}. Like other types of moments, Dixon's depend on a choice of origin. This origin is described by a reference worldline $\mathcal{Z}$, and introducing an arbitrary parameter $s$, we parameterize the points on that worldline by $z(s)$. Dixon's moments also depend on a choice of foliation for the worldtube, each leaf of which is a hypersurface $\Sigma(s) \ni z(s)$ formed from the intersection of all geodesics that pass through $z(s)$ and are orthogonal at that point to a given timelike vector $t^a(s)$. The reference worldline, one leaf of the foliation, and the orthogonal timelike vector are illustrated in Fig. \ref{Fig:worldtube}.

Various conditions have been proposed to fix $z(s)$ and $t^a(s)$ such that $\mathcal{Z}$ can be interpreted as a ``centroid'' or ``center of mass'' \cite{EhRu.77, Di.79, CoNa.15}. For example, the Tulczyjew-Dixon condition selects a worldline such that the linear momentum $p_a$ and the angular momentum $S^{ab} = S^{[ab]}$ satisfy
\begin{equation}
    p_a S^{ab} = 0.
    \label{centroid}
\end{equation}
This may be interpreted as demanding that the mass dipole moment vanish in the frame with vanishing 3-momentum. Simultaneously, each hypersurface $\Sigma(s)$ can be adapted to the vanishing 3-momentum frame by requiring $t^a(s)$ to be proportional to $p^a(s)$. Despite the highly-implicit nature of these conditions, existence and uniqueness results have been established under suitable conditions \cite{Scha.I.79, Scha.II.79}. Regardless, the majority of this paper does not assume any particular centroid conditions. 

Once a reference worldline and a foliation have been fixed (by whatever method), they can be used to introduce an infinite set of tensors
\begin{equation}
    \{ p_a(s), S^{ab}(s), J^{abcd}(s), J^{abcde}(s), \ldots\}    
\end{equation}
along $\mathcal{Z}$. The linear momentum $p_a$ and the angular momentum $S^{ab} = S^{[ab]}$ are associated with a body's monopole and dipole moments, respectively. For all $n \geq 0$, the rank $n+4$ tensors $J^{e_1 \cdots e_n abcd}$ instead describe a body's $2^{n+2}$-pole moments. These tensors are symmetric on their first $n$ indices and antisymmetric on both their final and their penultimate index pairs:
\begin{subequations}
\label{sym1}
\begin{align}
    J^{e_1 \cdots e_n abcd} &= J^{e_1 \cdots e_n [ab]cd} 
    \\
    &=  J^{e_1 \cdots e_n ab[cd]},
    \\
    &= J^{(e_1 \cdots e_n) abcd}.
\end{align}
\end{subequations}
Dixon's quadrupole and higher-order moments are additionally constrained by
\begin{subequations}
\label{sym2All}
\begin{align}
    J^{e_1 \cdots e_n a[bcd]} &= 0 ,
    \label{sym2a}
    \\
    J^{e_1 \cdots e_{n-1} [e_n ab]cd} &= 0 ,
    \label{sym2b}
    \\
    t_{e_1} J^{e_1 \cdots e_n abcd} &= 0,
    \label{sym3}
\end{align}
\end{subequations}
where the last two of these expressions hold only when $n \geq 1$. These are the only algebraic constraints on the multipole moments. As explained in Sec. 11 of \cite{Di.15}, it follows that $J^{e_1 \cdots e_n abcd}$ has $(n+4)(3n+5)$ independent components. A quadrupole moment $J^{abcd}$ thus depends on 20 independent components; an octupole moment $J^{eabcd}$ depends on 40.

While \eqref{sym1} and \eqref{sym2All} describe all \textit{algebraic} constraints on Dixon's moments, there are \textit{differential} constraints as well. These are entirely given by the Mathisson-Papapetrou-Dixon (MPD) equations
\begin{subequations}
\label{Dixon}
\begin{align}
	\dot{p}_a &= -\frac{1}{2} R_{abcd} \dot{z}^b S^{cd} + F_a, 
	\label{pDot}
	\\
	\dot{S}^{ab}  &= 2 p^{[a} \dot{z}^{b]} + N^{ab}.
	\label{Sdot}
\end{align}
\end{subequations}
The MPD equations describe how $p_a$ and $S^{ab}$ evolve along the reference worldline. The force $F_a$ and the torque $N^{ab} = N^{[ab]}$ that appear here can be determined from \eqref{genF} and \eqref{genFFN} below; see also \eqref{FNquad0} and \eqref{FNoct0} for the quadrupolar and octupolar contributions to these quantities. The MPD equations may be viewed as consequences of the conservation equation \eqref{stressCons}. While that equation constrains the evolution of a body's linear and angular momenta, it does not constrain the quadrupole and higher-order moments. Physically, the monopole and the dipole moments of $T^{ab}$ evolve ``universally;'' the higher-order moments do not. These latter moments can evolve differently for bodies with different compositions and different internal configurations. Two initially-identical objects may therefore fall differently, depending on their internal compositions.

If an object's stress-energy tensor is known, its moments can be computed at ``time'' $s$ by evaluating suitable integrals over $\Sigma(s)$; see Eq. (280) of \cite{Di.15}. If it is instead the moments that are known, the corresponding stress-energy tensor\footnote{If only a finite number of moments are nonzero, $T^{ab}$ is necessarily a distribution concentrated on $\mathcal{Z}$. A true extended body must have an infinite number of nonzero multipole moments.} can be found by evaluating Eq. (80) of that same reference, or Eq. (92) of \cite{Di.79}. Dixon's multipole moments therefore contain the same information as the stress-energy tensor $T^{ab}$. One can describe a body either by its stress-energy tensor, which is a tensor field with support on a spatially-compact worldtube, or by its multipole moments, which form a countably-infinite set of tensors along the reference worldline. What is special about Dixon's definitions is that they exactly convert stress-energy conservation---a \textit{partial} differential equation---into the MPD equations---which are a finite collection of \textit{ordinary} differential equations. Less sophisticated definitions for the multipole moments are typically constrained by an infinite number of ordinary differential equations. Dixon reduced this infinite set to a finite one without any loss of information. For that reason, he described his moments as ``reduced.''

\subsection{Generalized momentum and generalized force}

It is convenient here to utilize a reformulation of Dixon's formalism \cite{Ha.08, Ha.15} that views both the linear and the angular momenta as two components of the ``generalized momentum''
\begin{equation}
    \mathscr{P}_\xi (s) \equiv \int_{\Sigma(s)} \! \ud S_a T^{ab} \xi_b .
    \label{Pdef}
\end{equation}
Here, the vector fields $\xi^a$ are to be drawn from a certain ten-dimensional space of ``generalized Killing fields'' introduced in \cite{Ha.08}. For our purposes, the generalized Killing fields have three important properties. First, any genuine Killing field that may exist is also a generalized Killing field. Second, Killing's equation is always satisfied at least through first order on the reference worldline $\mathcal{Z}$, in the sense that
\begin{equation}
    \Lie_\xi g_{ab} \big|_{\mathcal{Z}} = \nabla_a \Lie_\xi g_{bc} \big|_{\mathcal{Z}} = 0.
    \label{genKilling}
\end{equation}
Third, $\xi^a$ is uniquely specified by its value and that of its first derivative anywhere on $\mathcal{Z}$. In fact, it is linear in that data. This last property implies that $\mathscr{P}_\xi(s)$ must be a linear combination of $\xi^a(z(s))$ and $\nabla_a \xi_b (z(s))$. The coefficients in that linear combination are identified as a body's linear momentum $p_a$ and angular momentum $S_{ab}$:
\begin{equation}
    \mathscr{P}_\xi = p_a \xi^a + \frac{1}{2} S^{ab} \nabla_a \xi_b. 
    \label{genPpS}
\end{equation}
The generalized momentum is useful because i) it allows both the linear and the angular momenta to be manipulated simultaneously, ii) the relevant objects are scalars rather than higher-rank tensors, and iii) it provides an immediate connection between conservation laws and symmetries.

This last point becomes apparent when considering the rate of change of the generalized momentum. Applying \eqref{stressCons}, the generalized force $\gF_\xi \equiv \ud \mathscr{P}_\xi / \ud s$ is seen to be given by
\begin{equation}
	\gF_\xi  = \frac{1}{2} \int_{\Sigma(s)} \! \ud S_c w^c T^{ab} \Lie_\xi g_{ab}   ,
    \label{genF}
\end{equation}
where $w^c$ is any time-evolution vector field for the foliation. If $\psi^a$ is Killing, $\gF_\psi$ vanishes and $\mathscr{P}_\psi = p_a \psi^a + \frac{1}{2} S^{ab} \nabla_a \psi_b$ is conserved. More generally, the generalized force measures the degree by which $\xi^a$ fails to be Killing. It is always possible to write $\gF_\xi(s)$ as a linear combination of $\xi^a (z(s))$ and $\nabla_a \xi_b (z(s))$, and the coefficients in that linear combination can be shown to be the force and the torque that appear in the MPD equations \eqref{Dixon}:
\begin{equation}
    \label{genFFN}
    \gF_\xi = F_a \xi^a + \frac{1}{2} N^{ab} \nabla_a \xi_b. 
\end{equation}
Note that the generalized force does not include the spin-curvature coupling $R_{abcd} \dot{z}^b S^{cd}$ in \eqref{pDot}, nor the $p^{[a} \dot{z}^{b]}$ term in \eqref{Sdot}. These terms are kinematic in origin and should not be regarded as ``proper'' forces or torques \cite{Di.79, Ha.15}. They arise when differentiating $\xi^a$ and $\nabla_a \xi_b$ in \eqref{genPpS} with respect to $s$ to find the generalized force.

Integral expressions for $F_a$ and $N^{ab}$ can be obtained by combining \eqref{genF}, \eqref{genFFN}, and an appropriate expression for the generalized Killing fields. However, this is rarely useful as-is. It is much more common to expand in a multipole series, which effectively assumes that $g_{ab}$ can be expanded in a Taylor series about $z(s)$. Doing so results in the formal series \cite{Di.74,Di.79,Ha.15},
\begin{equation}
	\gF_\xi = 2 \sum_{n=2}^\infty \frac{n-1}{(n+1)!} J^{c_1 \cdots c_{n-1} a c_n b} \Lie_\xi g_{ab, c_1 \cdots c_n},
    \label{genFexpandFull}
\end{equation}
where $g_{ab, c_1 \cdots c_n}$ denotes the $n$th tensor extension of the metric\footnote{The $n$th tensor extension $f_{\ldots, c_1 \cdots c_n}(x)$ of a tensor field $f_{\ldots}$, at a point $x$, is defined to be the unique tensor field that coincides with $n$ partial derivatives of $f_{\ldots}$ in a Riemann normal coordinate system with origin $x$ \cite{Di.74, Ve.23}.}. Forces and torques thus involve only the quadrupole and higher-order moments of $T^{ab}$. 

The multipole expansion in \eqref{genFexpandFull} is formal in two senses. First, the given series does not necessarily converge to $\gF_\xi$. It is expected only to be asymptotic and should be truncated at finite order. Second, and more subtly, the metric that appears in that series should not necessarily be interpreted as the physical metric. A body's own contribution to the spacetime metric varies considerably throughout any of its cross-sections, which invalidates the Taylor expansion here. An analogous problem appears in Newtonian gravity, and it is resolved there by expanding the external potential rather than the physical potential in all laws of motion \cite{Da.87, Ha.15, Di.79}. A similar procedure can be justified in general relativity \cite{Ha.12, Ha.15, Ha.25}, where portions of the self-field involving problematic short lengthscales can be shown not to contribute\footnote{Short-lengthscale portions of the self-field do actually contribute to $\mathscr{F}_\xi$, but only in a way that can be eliminated by replacing the stress-energy with a ``dressed'' or ``effective'' counterpart that takes into account the inertia of the self-field. All momenta here should technically be interpreted as derived from such an effective stress-energy tensor.} to $\mathscr{F}_\xi$. A nonlocal transformation applied to the physical metric results in a certain effective metric, and it is this effective metric that should be used in all laws of motion. While the effective potential in Newtonian gravity is interpreted purely as ``external,'' some of the effective metric in general relativity can involve aspects of an object's self-field.  Regardless, we ignore these details here and treat our objects as though they were extended test bodies moving in fixed background spacetimes. 

Our interest in this paper is on forces and torques at octupolar order, which corresponds to $n=3$ in the generalized force expansion \eqref{genFexpandFull}. The relevant tensor extensions then depend on the Riemann tensor via \cite{Di.74}
\begin{equation}
   	g_{ab,c d} = \frac{2}{3} R_{a(c d)b}, \qquad g_{ab,c d e} = \nabla_{(c} R_{|a| d e) b}.
\end{equation}
Repeatedly applying the Bianchi identities, it follows that the quadrupolar and octupolar generalized forces are given by
\begin{equation}
	\gF_\xi = - \frac{1}{6} J^{abcd} \Lie_\xi R_{abcd} - \frac{1}{12} J^{abcde} \Lie_\xi \nabla_a R_{bcde}+ \ldots
	\label{genFexpand}
\end{equation}
The force and the torque appearing in the MPD equations can be obtained from this by writing all Lie-derivatives in terms of covariant derivatives and comparing the result with \eqref{genFFN}. At quadrupolar order, the resulting expressions are well-known \cite{DiI.70}:
\begin{subequations}
    \label{FNquad0}
    \begin{align}
        F_{a}^{\rm quad} &= -\frac{1}{6} J^{bcde}\nabla_{a} R_{bcde}, \\
        N^{ab}_{\rm quad} &= -\frac{4}{3} J^{cde[a}R_{cde}^{\phantom{cde} b]}.
    \end{align}
\end{subequations}
The octupolar forces and torques
\begin{subequations}
    \label{FNoct0}
    \begin{align}
        F_a^{\rm oct} &= - \frac{1}{12} J^{bcdef} \nabla_a \nabla_b R_{cdef}, \\
        N^{ab}_{\rm oct} &= \frac{1}{3} J^{c d e f [a} \left( 2 \nabla_c R^{\phantom{def}b]}_{edf} + \nabla^{b]} R_{cfde} \right),
    \end{align}
\end{subequations}
are more obscure, but have been discussed before in App.~A of \cite{Ma.15}. Below, we largely prefer to work with generalized forces rather than separate forces and torques.

\subsection{Integral expressions for the multipole moments}

Some intuition for Dixon's moments can be gained by writing them in terms of a body's stress-energy tensor. A multipole moment at time $s$ then involves an integral over $\Sigma(s)$ that is linear in $T^{ab}$. Completely general expressions of this type may be found in Sec. 9 of \cite{Di.74} and in Sec. 24 of \cite{Di.15}, although they involve a variety of nontrivial bitensors that are difficult to evaluate. The multipolar integrals simplify considerably in flat spacetime and in terms of inertial coordinates\footnote{We use Greek indices to denote coordinates components that can run from 0 to 3. The Latin indices $i,j,k,l,m$ used below instead run only from 1 to 3. The  $a,b,\ldots$ used in the majority of this paper are abstract indices.} $x^\alpha$. In that context, the coordinate components of the quadrupole and the octupole moments reduce to
\begin{align}
    J^{\alpha \beta \gamma \delta} = \int_{\Sigma(s)} \! \ud S_\sigma w^\sigma (x-z)^{[\alpha} \Big[ T^{\beta][\delta} + ( \dot{z}^{\beta]} T^{\lambda [ \delta}
    \nonumber
    \\
    {} + T^{\beta]\lambda} \dot{z}^{[\delta} ) \nabla_\lambda \tau \Big] (x-z)^{\gamma]},
    \label{quadInt}
\end{align}
and
\begin{align}
    J^{\rho \alpha\beta\gamma\delta} = \int_{\Sigma(s)} \! \ud S_\sigma w^\sigma  (x-z)^\rho (x-z)^{[\alpha} \Big[ T^{\beta][\delta} 
    \nonumber
    \\
    {} + \frac{1}{2} ( \dot{z}^{\beta]} T^{\lambda [ \delta} + T^{\beta]\lambda} \dot{z}^{[\delta} ) \nabla_\lambda \tau \Big] (x-z)^{\gamma]} ,
    \label{octInt}
\end{align}
where $\tau$ is defined such that $\tau(x) = s$ for every $x \in \Sigma(s)$. Then, $w^\alpha \nabla_\alpha \tau = 1$. These expressions remain good approximations even in curved spacetimes if the object of interest is sufficiently small and the $x^\alpha$ are reinterpreted as Riemann normal coordinates with origin $z(s)$. 

Further simplifications arise in the case where the reference worldline is chosen such that $z^0(s) = s$ and $z^i(s) =0$, and where $\Sigma(s)$ corresponds to the $x^0 = s$ hypersurface. Then,
\begin{subequations}
    \label{quadFlat}
    \begin{align}
        J^{i 0 j 0} &= \frac{3}{4} \int \ud^3 x \, x^i x^j T^{00} ,
        \\
        J^{i j k 0} &= \int \ud^3 x \, x^k x^{[i} T^{j] 0} ,
        \\
        J^{ijkl} &= \int \ud^3 x \, x^{[i} T^{j][l} x^{k]}.
    \end{align}
\end{subequations}
The first line here is proportional to the full (as opposed to trace-free) Newtonian quadrupole moment of the energy density $T^{00}$. The second line involves a kind of momentum quadrupole, while the third line may be interpreted as being proportional to a stress quadrupole. A similar decomposition for the octupole moment results in
\begin{subequations}
    \label{octFlat}
    \begin{align}
        J^{m i 0 j 0} &= \frac{1}{2} \int \ud^3 x \, x^i x^j x^m T^{00}, 
        \\
        J^{m i j k 0} & = \frac{3}{4} \int \ud^3 x \, x^k x^m x^{[i} T^{j]0},
        \\
        J^{m i j k l } & = \int \ud^3 x\, x^m x^{[i} T^{j][l} x^{k]}.
    \end{align}
\end{subequations}
These components again have straightforward interpretations in terms of Newtonian-like octupole moments of $T^{00}$, $T^{i0}$, and $T^{ij}$. Further discussion on $3+1$ decompositions of the multipole moments is contained in Sec. \ref{Sec:31decomp} below.

\section{Motion is not affected by traces of the moments}
\label{Sec:NoTrace}

In Newtonian gravity, objects typically move in response to an external field that satisfies the vacuum field equation---the Laplace equation. This equation can be used to show that traces of an object's multipole moments cannot affect its motion \cite{PoWi, Ha.15, Th.80}. However, the trace-free components of the moments do not uniquely determine an object's mass distribution. It is therefore not possible to use the motion of a Newtonian body to completely infer its internal structure, at least without additional assumptions.

A similar situation has been noted before in Dixon's relativistic theory, at least at quadrupolar order \cite{Ha.10, BiGe.14, Ha.20}. To see this, first assume that the vacuum Einstein equation
\begin{equation}
    R_{ab} = \Lambda g_{ab}
    \label{Einstein}
\end{equation}
holds, perhaps with a cosmological constant $\Lambda$. Given any generalized Killing field $\xi^a$, it is then a consequence of \eqref{genKilling} that on the reference worldline $\mathcal{Z}$,
\begin{equation}
	g^{ac} \Lie_\xi R_{abcd} = \Lie_\xi R_{bd} = \Lambda \Lie_\xi g_{bd} = 0.
	\label{LieRiemannTrace}
\end{equation}
Separately, it follows from \eqref{genFexpand} that the quadrupolar generalized force is given by
\begin{equation}
    \gF_\xi^{\rm quad} \equiv - \frac{1}{6} J^{abcd} \Lie_\xi R_{abcd}    .
    \label{FgenQuad0}
\end{equation}
Combining these two equations shows that any transformation that shifts  $J^{abcd}$ by an outer product with the inverse metric must preserve the generalized force. For example, $\gF_\xi^{\rm quad}$ is preserved if the $J^{abcd}$ appearing on the right-hand side of \eqref{FgenQuad0} is replaced by the trace-free quadrupole\footnote{We typically use calligraphic fonts to denote trace-free components of multipole moments. Here, $\mathscr{J}^{abcd}$ is the trace-free component of $J^{abcd}$.}
\begin{align}
	\Jtf^{abcd} = J^{abcd} + g^{b[c} J^{d]fa}{}_{f}  - g^{a[c} J^{d]fb}{}_{f}
	\nonumber
	\\
	{} + \frac{1}{3} J^{fh}{}_{fh} g^{a[c} g^{d]b}.
	\label{quadTF}
\end{align}
The generalized force is also preserved if the Riemann tensor $R_{abcd}$ is replaced by the Weyl tensor $C_{abcd}$. Implementing both replacements shows that
\begin{equation}
    \gF_\xi^{\rm quad} = - \frac{1}{6} \Jtf^{abcd} \Lie_\xi C_{abcd}
    \label{FgenQuad}
\end{equation}
in any vacuum background. 

Trace-free quadrupoles are constrained by the same index symmetries as four-dimensional Weyl tensors, which are well-known to have ten independent components. There are therefore at least $20-10=10$ components of $J^{abcd}$ that cannot affect a body's motion in a vacuum background. Of course, it is not only $J^{abcd}$ and $\Jtf^{abcd}$ that result in the same forces and torques. An equivalence class is formed by all rank-4 tensors that are related to a given quadrupole moment by outer products with the inverse metric that also preserve the quadrupolar index symmetries. It is really this equivalence class that matters for motion in vacuum backgrounds.

We now extend these comments from quadrupolar to octupolar order. From \eqref{genFexpand}, first note that the octupolar generalized force is
\begin{equation}
    \gF_\xi^{\rm oct} \equiv - \frac{1}{12} J^{abcde} \Lie_\xi \nabla_a R_{bcde}. 
    \label{FgenOct0}
\end{equation}
It follows from \eqref{genKilling}, the vacuum Einstein equation \eqref{Einstein}, and the differential Bianchi identity that
\begin{equation}
    g^{ab} \Lie_\xi \nabla_a R_{bcde} = g^{ce} \Lie_\xi \nabla_a R_{bcde} = 0    
    \label{LieDRiemannTrace}
\end{equation}
on $\mathcal{Z}$. Any outer product with the inverse metric can thus be added to the octupole moment $J^{abcde}$ without affecting either the force or the torque. Moreover, it is exactly outer products of this kind that appear in the difference between the full octupole moment $J^{abcde}$ and its trace-free counterpart $\Jtf^{abcde}$. This is apparent from 
\begin{align}
	\Jtf^{abcde} = J^{abcde} + 2 ( j^{[de] [b} g^{c]a} + j^{[bc][d} g^{e]a} )
	\nonumber
	\\
	{} + j^{ae[b} g^{c]d}  - j^{a d [b} g^{c] e} ,
    \label{Projector4D}
\end{align}
which is a consequence of \eqref{Kdef} and \eqref{EfromL} in App. \ref{App:TF}. The tensor $j^{abc} = j^{a(bc)}$ that appears here is linear in $J^{abcde}$, and is essentially the $k^{abc}$ given by \eqref{Adef}. In this context, where the dimension $d$ is equal to 4,
\begin{align}
    j^{abc} \equiv \frac{1}{10} ( 6 J^{abdc}{}_{d} + 4 J^{(bc)da}{}_{d} -  J^{ade}{}_{de} g^{bc} ).
    \label{smallj}
\end{align}
Regardless, it follows immediately from \eqref{Einstein}, \eqref{LieDRiemannTrace}, and \eqref{Projector4D} that
\begin{equation}
    \gF_\xi^{\rm oct} = - \frac{1}{12} \Jtf^{abcde} \Lie_\xi \nabla_a C_{bcde}
    \label{FgenOct}
\end{equation}
in any vacuum spacetime. Only the trace-free component of an octupole moment can thus affect an object's motion. 

Besides being fully trace-free, $\Jtf^{abcde}$ satisfies analogs of the constraints \eqref{sym1}, \eqref{sym2a}, and \eqref{sym2b}:
\begin{subequations}
\label{Jtf1}
\begin{gather}
	\Jtf^{abcde} = \Jtf^{a[bc]de} = \Jtf^{abc[de]} ,
    \\
	\Jtf^{[abc]de} = \Jtf^{ab[cde]} = 0.
    \label{Jtf2}
\end{gather}
\end{subequations}
However, the orthogonality condition \eqref{sym3} is not preserved by the trace-free projection; it is not generically true that $t_a \Jtf^{abcde}$ vanishes. We show in Sec. \ref{Sec:31decomp} below that the trace-free octupole moment has 24 independent components, so there are at least $40-24= 16$ components of $J^{abcde}$ that cannot affect an object's motion in a vacuum background. 

As in the quadrupole case, it is really an equivalence class of octupole moments that is relevant for motion in vacuum backgrounds. Any rank-5 tensors that satisfy \eqref{Jtf1} while differing only by outer products with the inverse metric are in the same equivalence class. All such moments produce the same forces and torques.  

Interestingly, our argument that traces cannot affect motion does not extend straightforwardly beyond octupole order. At the next (hexadecapole) order, we need the tensor extension \cite{Ha.15}
\begin{equation}
	g_{ab,cdef} = \frac{6}{5} \nabla_{(c} \nabla_d R_{|a|ef)b} + \frac{16}{15} R_{a(cd}{}^{h} R_{|b|ef)h}
	\label{g4extend}
\end{equation}
in order to evaluate the generalized force \eqref{genFexpandFull}. A Lie derivative of this is to be contracted with the hexadecapole moment $J^{cdeafb}$, and one might ask if outer products with the inverse metric can be subtracted from that moment without affecting the motion. However, attempting to show this by analogy with \eqref{LieRiemannTrace} or \eqref{LieDRiemannTrace} fails. This is because, e.g.,
\begin{align}
    g^{ab} \Lie_\xi g_{ab,cdef} = \frac{16}{15} \Lie_\xi [ R_{a(cd}{}^h R^{a}{}_{ef) h} ] 
    \label{hexa}
\end{align}
in vacuum spacetimes, which does not necessarily vanish. While there are likely still components of the hexadecapole moment that cannot affect a body's motion when the vacuum Einstein equation is satisfied, they will require more care to isolate. 

Very recently, Amancio, Mosna, and Vieira~\cite{AmMoVi.26} have shown that for bodies satisfying a certain definition of spherical symmetry, the first nontrivial extended-body forces in vacuum backgrounds generically arise at hexadecapolar order. With their definitions, the multipole moments can be written entirely in terms of outer products of the metric and the momentum. Bodies that are spherically symmetric in that sense are therefore examples of objects where terms such as \eqref{hexa} can affect motion in vacuum backgrounds.

While the remainder of this paper limits itself only to vacuum spacetimes, and only to quadrupolar and octupolar effects, it can sometimes be interesting to consider non-vacuum backgrounds where $R_{ab} \neq \Lambda g_{ab}$. A system might, for example, contain strong electromagnetic fields, or perhaps dark matter that interacts only gravitationally and interpenetrates the body of interest. One might also consider theories of gravity other than general relativity. Regardless, various extended-body effects are possible in these cases: A spacecraft in a Friedmann-Lema\^{\i}tre-Robinson-Walker spacetime can control its moments in order to translate itself arbitrarily, even with a vanishing (and conserved) 3-momentum \cite{Ha.07}. Simply-pulsating spherically-symmetric stars can also move chaotically in spherically-symmetric backgrounds with integrable geodesics \cite{Mo.24, AmMoVi.24}. However, these effects are not unique features of non-vacuum backgrounds, at least at this level of description: Arbitrary translation with vanishing 3-momentum is also possible in vacuum Kasner backgrounds \cite{Ha.23}, and extended-body effects can introduce chaos in both Schwarzschild \cite{MoRoVi.22} and Kerr \cite{Ra.NoCarter.26} spacetimes (for bodies that are not spherically symmetric).

\section{Mass and momentum moments}
\label{Sec:31decomp}

Through octupolar order, forces and torques in vacuum backgrounds depend only on the trace-free components of the quadrupole and octupole moments. However, these moments can be difficult to work with due to their many indices and symmetries. In this section and the next, we provide simplifying decompositions for the trace-free quadrupole and octupole moments in terms of lower-rank tensors. The present section performs a $3+1$ split, using a timelike frame vector to decompose the moments into simpler, fully-spatial components. While the full moments can be written in terms of mass, momentum, and stress components, only mass and momentum components appear in the trace-free moments.

We begin by considering the quadrupole case, where a $3+1$ decomposition of $J^{abcd}$ was first given by Ehlers and Rudolph \cite{EhRu.77}. A similar decomposition of the trace-free quadrupole $\mathscr{J}^{abcd}$ was later obtained in \cite{Ha.20}, although few details were provided there. We present a more complete explanation of that result in Sec. \ref{Sec:31quad} while also expanding upon it. Similar decompositions for the full and the trace-free octupole moments are then obtained in Sec. \ref{sec:31oct}. 

Our decompositions require a choice of timelike frame vector. In the quadrupole case, any vector would result in the same presentation. In the octupole case, the orthogonality condition \eqref{sym3} makes it simpler to adopt the vector $t^a$ that is used to define the hypersurfaces $\Sigma(s)$. For notational simplicity, we nevertheless use $t^a$ to perform both decompositions. We also assume for simplicity that this vector has unit norm. Properties of the various quadrupole and octupole moments defined in this section are summarized in Sec. \ref{Sec:31summary} and in Table \ref{Table:Moments} below.  

\subsection{Decomposing the quadrupole}
\label{Sec:31quad}

In order to perform a $3+1$ decomposition of a body's trace-free quadrupole moment, we first review the decomposition of its full quadrupole moment into mass, momentum, and stress components. We then derive the force and the torque in terms these components, show how traces can be removed, and finally obtain a decomposition of the trace-free quadrupole moment in terms of trace-free mass and momentum components. 

\subsubsection{Full quadrupole moments}

Ehlers and Rudolph \cite{EhRu.77} have shown that given the symmetries \eqref{sym1} and \eqref{sym2a}, the full quadrupole moment must have the form\footnote{Our definitions differ slightly from those adopted by Ehlers and Rudolph \cite{EhRu.77}. The tensors $(M^{ab}, P^{abc}, S^{abcd})$ defined here are denoted there by $(m^{ab}, \frac{1}{2} \pi^{abc}, \tau^{abcd})$. Besides the differing symbols, note the factor of two difference between the momentum quadrupoles.}
\begin{align}
	J^{abcd} = S^{abcd} - 2 (t^{[a} P^{b]cd} + t^{[c} P^{d]ab}) - 3 t^{[a} M^{b][c} t^{d]} ,
    \label{quad31}
\end{align}
where $M^{ab} = M^{(ab)}$ is interpreted as the mass quadrupole, $P^{abc} = P^{a[bc]}$ as the momentum (or ``flow'' or ``current'') quadrupole, and $S^{abcd} = S^{[ab]cd} = S^{ab[cd]}$ as the stress quadrupole. These three tensors are fully orthogonal to $t^a$, and in addition,
\begin{equation}
    P^{[abc]} = 0 , \qquad S^{a[bcd]} = 0.
    \label{31quadCyclic}
\end{equation}
The mass, the momentum, and the stress quadrupoles contain 6, 8, and 6 independent components, respectively. Together, they determine the aforementioned 20 independent components of $J^{abcd}$.

Some intuition for this decomposition can be gained by determining how the various quadrupole components are related to a body's stress-energy tensor. Comparing \eqref{quadFlat} with \eqref{quad31} in the simple context where that former equation holds, the moments reduce to
\begin{subequations}
\label{quad31Integrals}
\begin{align}
    M^{ij} &= \int \ud^3 x \, x^i x^j T^{00}, 
    \\
    P^{ijk} &= \int \ud^3 x \, x^i x^{[j} T^{k]0},
    \\
    S^{ijkl} &= \int \ud^3 x \, x^{[i} T^{j][l} x^{k]},
\end{align}
\end{subequations}
in inertial coordinates. The integral for the mass quadrupole is exactly as one might expect. The momentum and the stress quadrupoles are perhaps nontrivial only in the antisymmetrizations that appear in the relevant integrals.

\subsubsection{Quadrupolar forces and torques}

In general, the stress quadrupole $S^{abcd}$ has the same symmetries as a three-dimensional Riemann tensor. Recalling that three-dimensional Weyl tensors always vanish, it must therefore be possible to write $S^{abcd}$ entirely in terms of its trace. More precisely, any stress quadrupole must satisfy
\begin{align}
    S^{abcd} = 2 ( h^{a[c} S^{d]eb}{}_{e} - h^{b[c} S^{d]ea}{}_{e} ) 
    \nonumber
    \\  
    {} - S^{ef}{}_{ef} h^{a[c} h^{d]b},
    \label{SquadTrace}
\end{align}
where
\begin{equation}
    h_{ab} \equiv g_{ab} + t_a t_b
    \label{hDef}
\end{equation}
denotes the spatial metric. Substituting this decomposition for $S^{abcd}$ into \eqref{quad31} while applying \eqref{LieRiemannTrace}, the quadrupolar generalized force \eqref{FgenQuad0} can be seen to reduce to
\begin{align}
    \gF_\xi^{\rm quad} = \frac{1}{6} t^a \big[ ( 3 M^{bc} + 4 S^{bec}{}_{e} ) t^d + 4 P^{bcd} \big] \Lie_\xi C_{abcd}
    \label{FgenQuad2}
\end{align}
in any vacuum background. 

Both $M^{ab}$ and $S^{acb}{}_{c}$ therefore contribute similarly to the motion. It also follows from \eqref{LieRiemannTrace} that any multiple of $h^{ab}$ can be added to these tensors without affecting $\gF_\xi^{\rm quad}$. The generalized force thus depends on the mass and the stress moments only via the trace-free ``mass'' quadrupole
\begin{equation}
   	\mathscr{M}^{ab} = \mathscr{M}^{(ab)} \equiv \left( M^{ab} + \tfrac{4}{3} S^{acb}{}_{c} \right)_\mathrm{TF}  .
    \label{MtfQuad}
\end{equation}
Here, the subscript ``TF'' denotes that $h_{ab}$ is to be used to remove any trace from the quantity in brackets. Also note that despite the name we have given it, $\mathscr{M}^{ab}$ is not purely a mass moment; it depends on both the full mass quadrupole $M^{ab}$ and the full stress quadrupole $S^{abcd}$. 

The full momentum quadrupole $P^{abc}$ can also be decomposed into trace and trace-free components, with the latter given by
\begin{equation}
    \mathscr{P}^{abc} = \mathscr{P}^{a[bc]} = P^{abc} + P_{d}{}^{d[b} h^{c]a}.
    \label{PtfQuad}
\end{equation}
Using this decomposition in \eqref{FgenQuad2} while again employing \eqref{LieRiemannTrace}, the quadrupolar generalized force reduces to
\begin{align}
    \gF_\xi^{\rm quad} = \frac{1}{6} t^a ( 3 \mathscr{M}^{bc} t^d + 4 \mathscr{P}^{bcd} ) \Lie_\xi C_{abcd}.
    \label{FgenQuad3}
\end{align}
This depends on the quadrupole moment only via $\mathscr{M}^{ab}$ and $\mathscr{P}^{abc}$, each of which has 5 independent components.

If desired, the quadrupolar force $F_a^{\rm quad}$ and the quadrupolar torque $N_{ab}^{\rm quad}$ can be extracted from $\mathscr{F}_\xi^\text{quad}$ by varying over all $\xi^a$ and $\nabla_a \xi_b$ in \eqref{genFFN}. Doing so,
\begin{subequations}
    \label{FNquad}
\begin{align}
    F_a^{\rm quad} &= \frac{1}{6} t^b ( 3 \mathscr{M}^{cd} t^e + 4 \mathscr{P}^{cde} ) \nabla_a C_{bcde},
    \\
     N^{ab}_{\rm quad} &= \frac{2}{3} \big[ 3 t^e ( t^c \mathscr{M}^{d[a} - \mathscr{M}^{cd} t^{[a}) 
    \nonumber
    \\
    & {}+ 2 ( 4 t^{(c} \mathscr{P}^{|e|d)[a}  -\mathscr{P}^{cde} t^{[a} ) \big] C_{dec}{}^{b]}.
\end{align}
\end{subequations}
These expressions are evaluated in Sec. \ref{Sec:Newtonian} below in nearly-Newtonian spacetimes.

\subsubsection{Trace-free quadrupole moments}

Although we now have expressions for the quadrupolar force and torque in terms of trace-free mass and momentum quadrupoles, it is sometimes useful to directly relate those moments to $\mathscr{J}^{abcd}$. This can be accomplished by comparing \eqref{FgenQuad} with \eqref{FgenQuad3} while applying appropriate symmetrizations in order to ensure compatibility with \eqref{sym1} and \eqref{sym2a}. Doing so results in 
\begin{align}
    \mathscr{J}^{abcd} = - \big[ 2 (t^{[a} \mathscr{P}^{b]cd} + t^{[c} \mathscr{P}^{d]ab}) + 3 t^{[a} \mathscr{M}^{b][c} t^{d]}   \big]_{\mathrm{TF}},
    \label{quadTF31}
\end{align}
where ``TF'' now denotes that all traces are to be removed using \eqref{quadTF}. This can also be written more explicitly as
\begin{align}
    \mathscr{J}^{abcd} =  -\big[ 2 (t^{[a} \mathscr{P}^{b]cd} + t^{[c} \mathscr{P}^{d]ab}) + 3 t^{[a} \mathscr{M}^{b][c} t^{d]} \big]
    \nonumber
    \\
     {}  + \frac{3}{4} ( g^{a[c} \mathscr{M}^{d]b} - g^{b[c} \mathscr{M}^{d]a} ).
\end{align}
Comparing with the analogous decomposition \eqref{quad31} of $J^{abcd}$, we observe that a $3+1$ decomposition of the trace-free quadrupole can be obtained by replacing $M^{ab}$ with $\mathscr{M}^{ab}$, replacing $P^{abc}$ with $\mathscr{P}^{abc}$, removing $S^{abcd}$, and then eliminating any remaining traces. No simpler result could have been expected. We show in Sec. \ref{sec:31oct} below that this simplicity does not extend to the octupole case.

The trace-free mass and momentum moments $\mathscr{M}^{ab}$ and $\mathscr{P}^{abc}$ encode exactly the same information as $\mathscr{J}^{abcd}$. That the former moments determine the latter follows immediately from \eqref{quadTF31}. That the latter moment also determines the former two can be seen by noting that
\begin{subequations}
\label{quadMomentProjections}
\begin{align}
    \mathscr{M}^{ab} &= \frac{8}{3} \mathscr{J}^{acbd} t_c t_d,
    \\
    \mathscr{P}^{abc} &= -\left[ \mathscr{J}^{bcad} t_d \right]_\perp ,
\end{align}
\end{subequations}
where ``$\perp$'' denotes that all free indices are to be projected using $h_{ab}$. 

A different decomposition for $\mathscr{J}^{abcd}$ was employed in \cite{Ha.20}, where $\mathscr{M}^{ab}$ was paired with\footnote{The tensors denoted here by $(P^{abc},\Pi^{ab})$ are equal to one half of the tensors denoted by $(\tilde{\Pi}^{abc},\Pi^{ab})$ in \cite{Ha.20}.} 
\begin{align}
	\Pi^{ab} = \Pi^{(ab)} \equiv -P^{(a}{}_{cd} \epsilon^{b)cde} t_e 
    \label{PiQuadDef}
\end{align}
rather than $\mathscr{P}^{abc}$ in an expression for $\mathscr{J}^{abcd}$. Either $\Pi^{ab}$ or $\mathscr{P}^{abc}$ can be used to parameterize the momentum contributions to $\mathscr{J}^{abcd}$. To see this, first note that \eqref{PtfQuad} can be used to show that the $P^{a}{}_{cd}$ appearing in the definition for $\Pi^{ab}$ may be replaced by $\mathscr{P}^{a}{}_{cd}$. Moreover,
\begin{equation}
    \mathscr{P}^{[a}{}_{cd} \epsilon^{b]cde} t_e = 0,
\end{equation}
which follows from contracting the left-hand side of this equation with $\epsilon_{abpq}$ while using that $\mathscr{P}^{abc}$ is trace-free. Hence,
\begin{equation}
    \Pi^{ab} = -\mathscr{P}^{a}{}_{cd} \epsilon^{bcde} t_e.
\end{equation}
Contracting this equation with $\epsilon_{bpqr} t^r$ then shows that
\begin{equation}
    \mathscr{P}^{abc} = - \frac{1}{2} \Pi^{ad} \epsilon_{d}{}^{bce} t_e.
    \label{Pitf}
\end{equation}
The tensors $\mathscr{P}^{abc}$ and $\Pi^{ab}$ thus encode the same information. Using $\Pi^{ab}$ to describe a momentum quadrupole has the advantage that, like $\mathscr{M}^{ab}$, it is symmetric and trace-free. The pair $(\mathscr{M}^{ab}, \Pi^{ab})$ might therefore appear to be more natural than $(\mathscr{M}^{ab}, \mathscr{P}^{abc})$. However, using $\mathscr{P}^{abc}$ in place of $\Pi^{ab}$ has the advantage that no Levi-Civita tensors appear in expressions for, e.g., the force or the torque.

The various quadrupole components considered here are summarized in Table \ref{Table:Moments} below. As a concrete illustration of this decomposition, mass and momentum components of the spin-induced quadrupole moment commonly used in the gravitational-wave literature \cite{Ma.15, Stei.11} are worked out in App.~\ref{App:inducedMoments}.
We also emphasize that the unit timelike vector $t^a$ appearing in our expressions is arbitrary. Despite the notation, it does not need to coincide with the vector used to define the hypersurfaces $\Sigma(s)$ in Sec. \ref{Sec:review}. 

\subsection{Decomposing the octupole}
\label{sec:31oct}

We now apply a similar analysis to decompose the full octupole moment into mass, momentum, and stress components. We then determine the octupolar force and torque, use Einstein's equation to remove traces, and write the trace-free octupole $\mathscr{J}^{abcde}$ in terms of trace-free mass and momentum components. Unlike in the quadrupole case, we now assume that the unit timelike vector $t^a$ used to perform our $3+1$ decompositions is also the vector used to fix the hypersurfaces $\Sigma(s)$.

\subsubsection{Full octupole moments}

Given \eqref{sym1} and \eqref{sym3}, any contraction of the full octupole moment $J^{abcde}$ with more than two copies of $t^a$ must vanish. It is therefore possible to write
\begin{align}
	J^{abcde} = S^{abcde} - \frac{3}{2}  ( t^{[b} P^{|a|c]de} + t^{[d} P^{|a|e]bc} )
	\nonumber
	\\
	{} - 2 t^{[b} M^{|a|c][d} t^{e]},
    \label{Jsplit}
\end{align}
where the mass octupole $M^{abc}$, the momentum octupole $P^{abcd}$, and the stress octupole $S^{abcde}$ are all orthogonal to $t^a$. Enforcing the symmetries \eqref{sym1}, \eqref{sym2a}, and \eqref{sym2b} shows that these moments must satisfy
\begin{subequations}
\begin{gather}
    M^{abc} = M^{(abc)}, \qquad P^{abcd} = P^{ab[cd]} = P^{(ab)cd},
    \label{MPsyms}
    \\
    S^{abcde} = S^{a[bc]de} = S^{abc[de]},
    \\
    P^{a[bcd]} = 0,
    \qquad 
    S^{ab[cde]} = S^{[abc]de} = 0.
    \label{tripleSym}
\end{gather}
\end{subequations}
Being symmetric on all three indices, the mass moment $M^{abc}$ has 10 independent components. Using that $P^{abcd}$ is symmetric on its first two indices, antisymmetric on its last two indices, and also that \eqref{tripleSym} provides three scalar constraints, the momentum octupole must have $6 \cdot 3 - 3 = 15$ independent components. The number of independent components of $S^{abcde}$ can be found by viewing it at as the same as three three-dimensional Riemann tensors restricted by the three scalar constraints provided by $S^{[abc]de} =0$. This again results in $3 \cdot 6 - 3 = 15$ independent components. Together, these counts recover the aforementioned $40 = 10+ 15 + 15$ independent components expected for $J^{abcde}$.

When the integral expressions \eqref{octFlat} for the coordinate components of $J^{abcde}$ hold, comparison with \eqref{Jsplit} shows that in the simplest flat-spacetime case, 
\begin{subequations}
\begin{align}
    M^{ijk}  &= \int \ud^3 x \, x^i x^j x^k T^{00},
    \\
    P^{ijkl} & = \int \ud^3 x \, x^i x^j x^{[k} T^{l]0},
    \\
    S^{ijklm} &= \int \ud^3 x \, x^i x^{[j} T^{k][m} x^{l]} ,
\end{align}
\end{subequations}
in inertial coordinates. These integrals are similar to their quadrupolar counterparts \eqref{quad31Integrals}. The coefficients in the expansion \eqref{Jsplit} were chosen in order to ensure that all integrals here have unit coefficients.

\subsubsection{Octupolar forces and torques}

In a general curved background, the mass, the momentum, and the stress octupoles can now be decomposed into components with and without traces. The simplest such decomposition involves the mass octupole, whose trace-free component is given by
\begin{equation}
    \tilde{\mathscr{M}}^{abc} = \tilde{\mathscr{M}}^{(abc)} = M^{abc} - \frac{3}{5} h^{(ab} M^{c)d}{}_{d} .
    \label{MtildeOct}
\end{equation}
This has $10-3 = 7$ independent components. The reason for the tilde here is made clear by \eqref{MtfOct} below, which defines a different trace-free mass octupole $\mathscr{M}^{abc}$. 

The trace-free component of the full momentum octupole $P^{abcd}$ can be shown to be given by
\begin{align}
    \mathscr{P}^{abcd} = P^{abcd} + \frac{2}{5} h^{ab} P^{e[cd]}{}_{e} + \frac{2}{15} (h^{a[c} P^{d]eb}{}_{e} 
    \nonumber
    \\
    \, \, {} + h^{b[c} P^{d]ea}{}_{e} ) - \frac{8}{15} ( P^{ea[c}{}_{e} h^{d]b} + P^{eb[c}{}_{e} h^{d]a}).
    \label{PtfOct}
\end{align}
This differs from the full momentum octupole by terms linear in $P^{aeb}{}_{e}$, which is trace-free and has $9-1=8$ independent components. The trace-free octupole $\mathscr{P}^{abcd}$ thus has $15-8 = 7$ independent components.

Moving on to the stress octupole, there are two types of trace that might be relevant: $S^{dabc}{}_{d}$ and $S^{abdc}{}_{d}$. In fact,  \eqref{tripleSym} implies that these two traces must be related via 
\begin{align}
    2S^{[ab]dc}{}_{d} = -  S^{dabc}{}_{d}.
    \label{StraceRelate}
\end{align}
All relevant information is therefore contained in the rank-3 stress trace
\begin{equation}
    S^{abc} = S^{a(bc)} \equiv S^{abdc}{}_{d}.
    \label{S3Def}
\end{equation}
Using \eqref{StraceRelate}, this is constrained to satisfy
\begin{equation}
    (S^{abc} - 2 S^{bac} )h_{bc} = 0,
\end{equation}
which comprises three scalar constraints. The stress trace $S^{abc}$ therefore contains $3 \cdot 6 - 3 = 15$ independent components. 

That count coincides with the number of independent components in $S^{abcde}$ itself. It suggests that---like the stress quadrupole---the stress octupole can be written entirely in terms of its trace. Indeed, the trace-free component of the stress octupole always vanishes. Applying the trace-free projector given in App. \ref{App:TF} in a three-dimensional setting with the metric $h_{ab}$, that observation implies that
\begin{align}
    S^{abcde} = {}& s^{a d [b} h^{c] e} - s^{ae[b} h^{c]d} 
	\nonumber
	\\
	& {} - 2 (s^{[de] [b} h^{c]a} + s^{[bc][d} h^{e]a}) ,
    \label{SoctDecomp}
\end{align}
where
\begin{align}
    s^{abc} \equiv \mathscr{S}^{abc} + \mathscr{S}^{(bc)a}  + \frac{1}{10} h^{bc} S^{ad}{}_{d}.
    \label{little_s}
\end{align}
This depends on the trace-free component $\mathscr{S}^{abc}$ of $S^{abc}$, which is given by
\begin{equation}
    \mathscr{S}^{abc} = \mathscr{S}^{a(bc)} = S^{abc} - \frac{1}{10} \big( 3 h^{bc} S^{ad}{}_{d} + h^{a(b} S^{c)d}{}_{d} \big) .
    \label{StfOct}
\end{equation}
The freedom \eqref{kGauge} has also been used here to remove any $h^{a(b} S^{c)d}{}_{d}$ term from $s^{abc}$. 

It is now possible to determine how the mass, the momentum, and the stress octupoles appear in the generalized force. Substituting \eqref{Jsplit} into \eqref{FgenOct0} first shows that
\begin{align}
    \gF_\xi^{\rm oct} = \frac{1}{12} ( 2  M^{acd} t^b t^e + 3 P^{acde} t^b  - S^{abcde} )
    \nonumber \label{OctForce1}
    \\
    {} \times \Lie_\xi \nabla_a C_{bcde}
\end{align}
in any vacuum background. Recalling \eqref{LieDRiemannTrace} while applying the trace decompositions \eqref{MtildeOct}, \eqref{PtfOct}, \eqref{SoctDecomp}, and \eqref{StfOct}, this expression can be rewritten as
\begin{align}
    \gF_\xi^{\rm oct} = \frac{1}{12} t^b \big[2 \mathscr{P}^{cd} t^a t^e + 2 ( \tilde{\mathscr{M}}^{acd} + 2 \mathscr{S}^{acd} ) t^e
    \nonumber
    \\
    {}+ 3 \mathscr{P}^{acde} \big]\Lie_\xi \nabla_a C_{bcde},
\end{align}
where 
\begin{equation}
    \mathscr{P}^{ab} = \mathscr{P}^{(ab)} \equiv P^{c (ab)}{}_{c}
    \label{PtfOct2}
\end{equation}
is automatically trace-free. Further defining 
\begin{equation}
    \mathscr{M}^{abc} = \mathscr{M}^{a(bc)} \equiv \tilde{\mathscr{M}}^{abc} + 2 \mathscr{S}^{abc},
    \label{MtfOct}
\end{equation}
the octupolar generalized force reduces to
\begin{align}
    \gF_\xi^{\rm oct} = \frac{1}{12} t^b \big(2 \mathscr{P}^{cd} t^a t^e + 2 \mathscr{M}^{acd} t^e + 3 \mathscr{P}^{acde} \big)
    \nonumber
    \\
    {} \times \Lie_\xi \nabla_a C_{bcde}
    \label{FgenOct2}
\end{align}
in any vacuum spacetime. Applying \eqref{genFFN} now allows one to extract the octupolar force
\begin{align}
    F_a^{\rm oct} = \frac{1}{12} t^c \big(2 \mathscr{P}^{de} t^b t^f + 2 \mathscr{M}^{bde} t^f + 3 \mathscr{P}^{bdef} \big)
    \nonumber
    \\
    {} \times \nabla_a \nabla_b C_{cdef},
\end{align}
and the octupolar torque 
\begin{align}
    N_{ab}^{\rm oct} = \frac{1}{6} t^f \big[ 2 (3 t^{c} \mathscr{P}^{de} -t^d \mathscr{P}^{ce}) t_{[a} - 4 t^c t^d \mathscr{P}^{e}{}_{[a} 
    \nonumber
    \\
    {} + 4 ( \mathscr{M}^{cde} t_{[a} - t^d \mathscr{M}^{ce}{}_{[a} - \mathscr{M}_{[a}{}^{e[c} t^{d]}) 
    \nonumber
    \\
    {} + 3 (\mathscr{P}^{cdef} t_{[a} + t^c \mathscr{P}_{[a}{}^{def} - 2 t^d \mathscr{P}_{[a}{}^{cef} 
    \nonumber
    \\
    {} + 2 t^f \mathscr{P}^{ced}{}_{[a} ) \big] \nabla_{|c|} C_{b]def}.
\end{align}
These expressions depend only on the trace-free mass\footnote{Recalling \eqref{MtfQuad} and \eqref{MtfOct}, the trace-free ``mass'' quadrupole $\mathscr{M}^{ab}$ and its octupolar counterpart $\mathscr{M}^{abc}$ both depend on the full mass moments as well as on the full stress moments. However, while the contributions of $M^{ab}$ and $S^{abcd}$ to $\mathscr{M}^{ab}$ are entirely degenerate, the contributions of $M^{abc}$ and $S^{abcde}$ to $\mathscr{M}^{abc}$ are not. This can be seen by noting that, e.g., $\mathscr{M}^{[ab]c}$ depends \textit{only} on the stress octupole.} octupole $\mathscr{M}^{abc}$ and the trace-free momentum octupoles $\mathscr{P}^{ab}$ and $\mathscr{P}^{abcd}$. They are computed in App.~\ref{App:inducedMoments} for the spin-induced octupole moment of \cite{Ma.15}.

More generally, although our $3+1$ decomposition for the trace-free quadrupole moment involves only a single trace-free momentum component, we have found it convenient to define \textit{two} trace-free momentum octupoles: $\mathscr{P}^{ab}$ and $\mathscr{P}^{abcd}$. These momentum octupoles contain 5 and 7 independent components, respectively, the sum of which coincides with the 12 independent components of $\mathscr{M}^{abc}$. One might therefore suspect that both momentum octupoles can be encoded in a single rank-3 tensor with the same structure as $\mathscr{M}^{abc}$. This is indeed the case, and the appropriate analog of the quadrupolar $\Pi^{ab}$ defined by \eqref{PiQuadDef} is
\begin{equation}
    \Pi^{abc} = \Pi^{a(bc)} \equiv \big( \mathscr{P}^{bc}{}_{df} + \mathscr{P}^{(b}{}_{d} h^{c)}{}_f \big) \epsilon^{adfe} t_e.
    \label{PiOctDef}
\end{equation}
This is fully trace-free. Furthermore, $\mathscr{P}^{ab}$ and $\mathscr{P}^{abcd}$ can be extracted from it using
\begin{subequations}
\begin{align}
    \mathscr{P}^{ab} &= -\frac{2}{3} \Pi^{[cd]a} \epsilon^{b}{}_{cde} t^e,
    \\
    \mathscr{P}^{abcd} &= \frac{1}{2} \Pi^{(abe)} \epsilon^{cd}{}_{ef}t^f.
\end{align}
\end{subequations}
Octupolar forces and torques may be described either in terms of the pair $(\mathscr{M}^{abc}, \Pi^{abc})$ or the triple $(\mathscr{M}^{abc}, \mathscr{P}^{ab}, \mathscr{P}^{abcd})$. Although the former representation is simpler in isolation, we employ the latter below in order to avoid a proliferation of Levi-Civita tensors.

\subsubsection{Trace-free octupole moments}

The octupolar generalized force \eqref{FgenOct2} can now be compared with \eqref{FgenOct} in order to write the trace-free octupole moment $\mathscr{J}^{abcde}$ in terms of $\mathscr{M}^{abc}$, $\mathscr{P}^{ab}$, and $\mathscr{P}^{abcd}$. However, the coefficients of $\Lie_\xi \nabla_a C_{bcde}$ in the former equation must first be arranged so they have the same index symmetries as $\mathscr{J}^{abcde}$. For example, the terms involving the momentum moments can be replaced with
\begin{subequations}
\label{Pterms}
\begin{align}
    t^a t^b t^e \mathscr{P}^{cd} &\mapsto t^a t^{[b} \mathscr{P}^{c][d} t^{e]},
    \\
    t^b \mathscr{P}^{acde} &\mapsto \frac{1}{2} ( t^{[b} \mathscr{P}^{c]ade} + t^{[d} \mathscr{P}^{e]abc}),
\end{align}
\end{subequations}
without affecting $\mathscr{F}_\xi^\text{oct}$. Like $\mathscr{J}^{abcde}$, these replacements are antisymmetric in $bc$ and $de$. They also vanish when antisymmetrized over either $abc$ or $cde$. Writing down a similar replacement for the $t^b t^e \mathscr{M}^{acd}$ term in \eqref{FgenOct2} can be accomplished by first using $\nabla_{[a} C_{bc]de} = 0$ to note that
\begin{align}
    t^b t^e \mathscr{M}^{acd} \Lie_\xi \nabla_a C_{bcde} = \frac{1}{2} ( 2 t^b \mathscr{M}^{(ac)d} + t^a \mathscr{M}^{[bc]d} )
    \nonumber
    \\
    {} \times t^e \Lie_\xi \nabla_a C_{bcde}.
\end{align}
Antisymmetrizing the terms on the right-hand side that are contracted with $\Lie_\xi \nabla_a C_{bcde}$ over $bc$ and $de$ while additionally symmetrizing over the interchange of these two pairs results in the replacement
\begin{align}
    t^b t^e \mathscr{M}^{acd} &\mapsto \frac{1}{4} \big[ t^{[d} \mathscr{M}^{e]a[b} t^{c]}  + t^{[b} (2 \mathscr{M}^{|a|c][d} 
    \nonumber
    \\
    &{} + \mathscr{M}^{c]a[d})t^{e]}  + t^a ( \mathscr{M}^{[bc][d} t^{e]} + \mathscr{M}^{[de][b} t^{c]} ) \big],
    \label{Mterms}
\end{align}
which does not affect $\mathscr{F}_\xi^\text{oct}$. The terms on the right-hand side here are again antisymmetric over $bc$ and $de$. They also vanish when antisymmetrized over either $abc$ or $cde$. 

Applying the replacements \eqref{Pterms} and \eqref{Mterms} to \eqref{FgenOct2}, comparison with \eqref{FgenOct} shows that
\begin{align}
    \mathscr{J}^{abcde} = - \frac{1}{2} \big[ 4t^a t^{[b} \mathscr{P}^{c][d} t^{e]}+ t^{[d} \mathscr{M}^{e]a[b}t^{c]}  + t^{[b} (2 \mathscr{M}^{|a|c][d} 
    \nonumber
    \\
    {}   + \mathscr{M}^{c]a[d} ) t^{e]}  + t^a (\mathscr{M}^{[bc][d} t^{e]} + \mathscr{M}^{[de][b} t^{c]} )
    \nonumber
    \\
    {}  + 3 (t^{[b} \mathscr{P}^{c]ade} + t^{[d} \mathscr{P}^{e]abc})\big]_{\rm TF},
    \label{JsplitTF}
\end{align}
where the ``TF'' subscript denotes that all traces are to be removed using \eqref{Projector4D}. Not only is the trace-free octupole entirely determined by $\mathscr{M}^{abc}$, $\mathscr{P}^{ab}$, and $\mathscr{P}^{abcd}$; the opposite is true as well. This can be verified by noting that the latter tensors are given by
\begin{subequations}
\label{octMomentProjections}
\begin{align}
    \mathscr{P}^{ab} &= \frac{20}{3} \mathscr{J}^{cdabe} t_c t_d t_e,
    \\
    \mathscr{M}^{abc} &= -\frac{4}{3} \big[ (3 \mathscr{J}^{(a|d|bc)e} + 5 \mathscr{J}^{da(bc)e})t_d t_e \big]_\perp,
    \\
    \mathscr{P}^{abcd} &=  \frac{4}{9} \big\{ \big[ (h^{a[c} \mathscr{J}^{|pq| d]b e} + h^{b[c} \mathscr{J}^{|pq| d]a e} )t_p t_q 
    \nonumber
    \\
    & \qquad \qquad \qquad \qquad 
    {}- 3 \mathscr{J}^{(a|cd|b)e} \big] t_e  \big\}_\perp ,
\end{align}
\end{subequations}
where the $\perp$ subscript again denotes that all free indices are to have temporal components removed using $h_{ab}$. Unlike in the quadrupole case, note that our $3+1$ expansion \eqref{JsplitTF} for $\mathscr{J}^{abcde}$ does not arise by applying a simple replacement rule to the expansion \eqref{Jsplit} for $J^{abcde}$. 

It can also be noted that although the full octupole moment is constrained by the orthogonality constraint $t_a J^{abcde} = 0$, combining \eqref{JsplitTF} with \eqref{Projector4D} and \eqref{smallj} shows that the trace-free octupole moment instead satisfies
\begin{align}
    t_a \mathscr{J}^{abcde} = \frac{1}{10} \big[ 12 t^{[b} \mathscr{P}^{c][d} t^{e]} + 3( \mathscr{P}^{e[b} g^{c]d} - \mathscr{P}^{d[b} g^{c]e} ) 
    \nonumber
    \\
    {} + 4 ( \mathscr{M}^{[bc][d} t^{e]} + \mathscr{M}^{[de][b} t^{c]} ) \big]. 
\end{align}
This is generically nonzero. However, even more can be said: We argue in Sec. \ref{Sec:NullDecomp} below that
the orthogonality constraint has no effect whatsoever on the space of trace-free octupole moments. As a consequence, the $3+1$ decomposition \eqref{JsplitTF} for $\mathscr{J}^{abcde}$ is valid for \textit{any} unit timelike $t^a$. However, if the $t^a$ used for that decomposition is not the one used to construct the leaves $\Sigma(s)$ of the foliation, the $3+1$ expansion for the full octupole would not be given by \eqref{Jsplit}. Relations between the full and the trace-free moments would also be considerably more complicated than the ones obtained above. 

\subsection{Summarizing the $3+1$ decompositions}
\label{Sec:31summary}

We have now defined a number of different tensors in our decompositions of Dixon's quadrupole and octupole moments. These tensors, as well as decompositions appearing in Sec. \ref{Sec:NullDecomp} below, are summarized in Table \ref{Table:Moments}. 

\begin{table*}
  \centering
  \setlength{\tabcolsep}{9pt}

  \begin{tabular}{l p{0.28\textwidth} l l l}
    \hline\hline
    Moment & Description & Vanishing permutations & Components & Reference \\[.3ex]
    \hline
    $J^{abcd}$, $\mathscr{J}^{abcd}$	&	Full \& TF quadrupoles of $T^{ab}$ & $(ab)$, $(cd)$, $[abc]$ & 20, 10 & \eqref{quadInt},  \eqref{quadTF}
    \\
    $M^{ab}$, $\mathscr{M}^{ab}$	& Full \& TF mass quadrupoles  & $[ab]$ & 6, 5 & \eqref{quad31}, \eqref{MtfQuad}
    \\
    $P^{abc}$, $\mathscr{P}^{abc}$ & Full \& TF momentum quadrupoles & $(bc)$, $[abc]$ & 8, 5 & \eqref{quad31}, \eqref{PtfQuad}
    \\
    $\Pi^{ab}$ & Alt. TF momentum quadrupole & $[ab]$ & 5 & \eqref{PiQuadDef}
    \\
    $S^{abcd}$  & Stress quadrupole & $(ab)$, $(cd)$, $[bcd]$ & 6 & \eqref{quad31}
    \\
    $J_0, \ldots, J_4$ & Quadrupole scalars & - & 10 total & \eqref{Ji}, \eqref{quad}
    \\[.2ex]
    \hline
    $J^{abcde}$, $\mathscr{J}^{abcde}$ & Full \& TF octupoles of $T^{ab}$ &	$(bc)$, $(de)$, $[abc]$, $[cde]$ & 40, 24 & \eqref{octInt},  \eqref{Projector4D}
    \\
    $M^{abc}$   & Full mass octupole & $[ab]$, $[bc]$ & 10 & \eqref{Jsplit}
    \\
    $\mathscr{M}^{abc}$ & TF mass octupole & $[bc]$ & 12 & \eqref{MtfOct}
    \\
    $P^{abcd}$, $\mathscr{P}^{abcd}$ & Full \& 1st TF momentum octupoles & $[ab]$, $(cd)$, $[bcd]$ & 15, 7 &  \eqref{Jsplit}, \eqref{PtfOct}
    \\
    $\mathscr{P}^{ab}$ & 2nd TF momentum octupole & $[ab]$ & 5 & \eqref{PtfOct2}
    \\
    $\Pi^{abc}$ & Alt. TF momentum octupole & $[bc]$ & 12 & \eqref{PiOctDef}
    \\
    $S^{abcde}$ & Full stress octupole & $(bc)$, $(de)$, $[abc]$, $[cde]$ & 15 & \eqref{Jsplit}
    \\
    $J_0^a, \ldots, J_4^a$ & Octupole vectors & - & 24 total & \eqref{Jioct}, \eqref{oct}
    \\[.2ex]
    \hline\hline
  \end{tabular}
  
  \caption{Summary of different quadrupole and octupole moments. The first group of symbols are all quadrupole moments while the second are all octupole moments. Calligraphic fonts are used for trace-free moments. Compared with analogous quadrupoles, octupoles are typically denoted with the same basic symbol but one additional index. The ``Vanishing permutations'' column lists the index permutations that must vanish (e.g., $J^{(ab)cd} = 0$). There can, however, be additional constraints that are not indicated here, such as orthogonality or trace conditions. The ``Components'' column lists the number of independent real components with all constraints taken into account. Where more than one symbol is listed on a given row, the order of those symbols in the leftmost column is the same as the ordering of their component counts. }
  \label{Table:Moments}
\end{table*}

 We have noted that the full quadrupole and octupole moments can be decomposed into mass, momentum, and stress components via \eqref{quad31} and \eqref{Jsplit}. Schematically, those results express the equivalencies
\begin{subequations}
\begin{align}
    J^{abcd} &\leftrightarrow (M^{ab},P^{abc}, S^{abcd}),
    \\
    J^{abcde} &\leftrightarrow (M^{abc},P^{abcd}, S^{abcde}),
\end{align}
\end{subequations}
where the moments appearing on the right-hand sides of these expressions are fully orthogonal to $t^a$ and can have nonzero traces.

More interesting are the $3+1$ decompositions for the trace-free quadrupole and octupole moments, which are given by \eqref{quadTF31} and \eqref{JsplitTF}. Those expressions depend only on certain trace-free mass and momentum moments:
\begin{subequations}
\label{JTFequiv}
\begin{gather}
    \mathscr{J}^{abcd} \leftrightarrow (\mathscr{M}^{ab}, \mathscr{P}^{abc}) \leftrightarrow (\mathscr{M}^{ab}, \Pi^{ab}), 
    \\
    \mathscr{J}^{abcde} \leftrightarrow (\mathscr{M}^{abc}, \mathscr{P}^{ab}, \mathscr{P}^{abcd}) \leftrightarrow (\mathscr{M}^{abc}, \Pi^{abc}).
\end{gather}
\end{subequations}
Two types of momentum moments are employed here, denoted by $\mathscr{P}^{\cdots}$ and $\Pi^{\cdots}$. While there is only one $\mathscr{P}$-type quadrupole, there are two such octupoles. There is instead only one $\Pi$-type quadrupole and only one $\Pi$-type octupole. Both types of momentum moment are, however, equivalent:
\begin{subequations}
\begin{gather}
    \mathscr{P}^{abc} \leftrightarrow \Pi^{ab},
    \\
    (\mathscr{P}^{ab}, \mathscr{P}^{abcd} ) \leftrightarrow \Pi^{abc}.
\end{gather}
\end{subequations}
The $\Pi$-type moments are convenient in that they have the same symmetries as the corresponding mass moments. The $\mathscr{P}$-type moments have the advantage that expressions involving them typically avoid the presence of Levi-Civita tensors. See, e.g., \eqref{FgenQuad3} and \eqref{FgenOct2} for the quadrupolar and octupolar generalized forces in terms of the $\mathscr{P}$-type moments. In the quadrupole case, there are 5 trace-free mass moments and 5 trace-free current moments. In the octupole case, there are 12 trace-free mass moments and 12 trace-free current moments. 

The following section provides additional decompositions for the trace-free quadrupole and octupole moments, employing a null tetrad rather than a timelike vector. This results in the equivalencies
\begin{subequations}
\label{JTFequiv2}
\begin{gather}
    \mathscr{J}^{abcd} \leftrightarrow (J_0, J_1, J_2, J_3, J_4), 
    \\
    \mathscr{J}^{abcde} \leftrightarrow (J_0^a, J_2^a, J_4^a), 
\end{gather}    
\end{subequations}
where the scalars $J_{\ldots}$ and the vectors $J^a_{\ldots}$ can all be complex. These decompositions may be viewed as supplementing the ones in \eqref{JTFequiv}.

\section{Null decomposition}
\label{Sec:NullDecomp}

The $3+1$ decomposition described in Sec.~\ref{Sec:31decomp} can be an intuitive way to describe an object's multipole moments in terms of quasi-Newtonian concepts. However, if the timelike frame vector used there is not adapted to the structure of the background spacetime, it may result in relatively complicated force and torque expressions. This complication can obscure precisely which portions of a multipole moment affect the motion and which do not. It can also make it difficult to determine if the effects of different multipolar components are degenerate, which would prevent them from being distinguished by observing an object's motion. Such questions are more easily addressed by expanding the moments in a basis that has been adapted to properties of the background spacetime rather than properties of the body itself. 

To this end, it has been found useful---at least in the quadrupole case---to perform decompositions that involve a null basis adapted to the principal null directions of the background curvature \cite{Ha.20, Ha.23}. Such a ``null'' decomposition has been used to show that, depending on the Petrov type of the spacetime, certain components of $\mathscr{J}^{abcd}$ cannot affect a body's motion. It has also been found that in some cases, certain torques are impossible and some forces are possible only in combination with appropriate torques. Similar results also arise in Newtonian gravity, where the analog of the Petrov classification involves classifying the eigenvectors of the tidal tensor \cite{Ha.21, Ha.23}. We nevertheless restrict ourselves here only to general relativity, reviewing the null decomposition in the quadrupole case and then extending it to also describe trace-free octupole moments. Our null decompositions allow trace-free quadrupoles to be written in terms of five complex scalars and trace-free octupoles to be written in terms of three complex vectors.

\subsection{Vector and bivector bases}

The null decomposition is based on the choice of a null tetrad $(\ell^a, n^a, m^a, \bar{m}^a)$, where $\ell^a$ and $n^a$ are real, $m^a$ and $\bar{m}^a$ are complex conjugates of one another, and the only non-vanishing inner products are
\begin{equation}
    \ell^a n_a =- 1, \qquad m^a \bar{m}_a = 1 .
    \label{orthonlm}
\end{equation}
In Sec. \ref{Sec:TypeD} below, $\ell^a$ and $n^a$ are aligned with the principal null directions of a type D spacetime. However, both the (vacuum) background and the tetrad can be viewed as arbitrary in this section. 

The null tetrad is employed here mainly in order to construct a complex basis of bivectors
\begin{equation}
    (X^{ab}, Y^{ab}, Z^{ab}, \bar{X}^{ab}, \bar{Y}^{ab}, \bar{Z}^{ab})    
\end{equation}
that can be used to expand antisymmetric index pairs arising either in an object's multipole moments or in the Weyl tensor. Relatively simple inner products between the various basis elements result from the definitions
\begin{subequations}
\label{defXYZ}
\begin{gather} 
    X^{ab} \equiv 2 \ell^{[a} m^{b]}, \qquad Y^{ab} \equiv 2 n^{[a} \bar{m}^{b]},
    \\
    Z^{ab} \equiv 2 (\ell^{[a} n^{b]} - m^{[a} \bar{m}^{b]}).
\end{gather}
\end{subequations}
For example, the only non-vanishing scalar contractions are  
\begin{equation} \label{prodXYZ}
    X^{ab}Y_{ab}=-2 , \qquad Z^{ab}Z_{ab}=-4,
\end{equation}
and their complex conjugates. If one pair of indices is contracted rather than two, we have that
\begin{equation}
\label{bivectorProds}    
\begin{aligned}
  X^{a c} Y^{b}{}_{c}  &= \frac{1}{2} (Z^{a b} - g^{a b}), & X^{a c} Z^{b}{}_{c} &= -X^{ab},
  \\
  Y^{a c} Z^{b}{}_{c} &= Y^{a b}, &
  Z^{a c} Z^{b}{}_{c} &= - g^{a b},
  \\
  X^{ac} \bar{X}^{b}{}_{c} &= \ell^a \ell^b, & X^{ac} \bar{Y}^{b}{}_{c} &= - m^a m^b ,  
    \\
    X^{ac} \bar{Z}^{b}{}_{c} &= 2 \ell^{(a} m^{b)}, & Y^{ac} \bar{Y}^{b}{}_{c} &= n^a n^b, 
    \\
     Y^{ac} \bar{Z}^{b}{}_{c} &= -2 n^{(a} \bar{m}^{b)}, & Z^{ac} \bar{Z}^{b}{}_{c} &= g^{ab} + 4  \ell^{(a} n^{b)} .
    \end{aligned}
\end{equation}
Working with a complex basis such as this effectively allows us to halve the number of parameters that are needed to describe a (real) multipole moment or a Weyl tensor.

\subsection{Decomposing the quadrupole}

It was observed in \cite{Ha.20} that since the trace-free quadrupole moment $\mathscr{J}^{abcd}$ has the same properties as a four-dimensional Weyl tensor, it is possible to define five complex ``quadrupole scalars'' $J_0, \ldots, J_4$ that are analogous to the five Weyl scalars $\Psi_0, \ldots, \Psi_4$ commonly used to describe a Weyl tensor $C_{abcd}$ \cite{Cha, SKMHH}. These latter scalars depend on the choice of tetrad, being defined by
\begin{subequations}
\label{PsiDef}
\begin{align}
	\Psi_0 &\equiv \frac{1}{4} C_{abcd} X^{ab} X^{cd}, \quad \Psi_1 \equiv \frac{1}{8} C_{abcd} X^{ab} Z^{cd},
	\\
	\Psi_2 &\equiv -\frac{1}{4} C_{abcd} X^{ab} Y^{cd} = \frac{1}{16} C_{abcd} Z^{ab} Z^{cd},
	\\
	\Psi_3 &\equiv -\frac{1}{8} C_{abcd} Y^{ab} Z^{cd}, \quad \Psi_4 \equiv \frac{1}{4} C_{abcd} Y^{ab} Y^{cd}.
\end{align}
\end{subequations}
In terms of them,
\begin{align}
	C_{abcd} = 2 \Re \big[ \Psi_0 Y_{ab} Y_{cd} + \Psi_1 (Y_{ab} Z_{cd} + Z_{ab} Y_{cd}) 
	\nonumber
	\\
	{} + \Psi_2 (Z_{ab} Z_{cd} - X_{ab} Y_{cd} - Y_{ab} X_{cd} )
	\nonumber
	\\
	{} -  \Psi_3 ( X_{ab} Z_{cd} + Z_{ab} X_{cd}) + \Psi_4 X_{ab} X_{cd} \big].
    \label{Weyl}
\end{align}
Similarly, five quadrupole scalars can be defined via
\begin{subequations}
\label{Ji}
\begin{align}
	J_0 &\equiv \frac{1}{4} \mathscr{J}_{abcd} X^{ab} X^{cd}, \quad J_1 \equiv \frac{1}{8} \mathscr{J}_{abcd} X^{ab} Z^{cd},
	\\
	J_2 &\equiv -\frac{1}{4} \mathscr{J}_{abcd} X^{ab} Y^{cd} = \frac{1}{16} \mathscr{J}_{abcd} Z^{ab} Z^{cd},
	\\
	J_3 &\equiv -\frac{1}{8} \mathscr{J}_{abcd} Y^{ab} Z^{cd}, \quad J_4 \equiv \frac{1}{4} \mathscr{J}_{abcd} Y^{ab} Y^{cd},
\end{align}
\end{subequations}
in which case
\begin{align}
	\mathscr{J}^{abcd} = 2 \Re \big[ J_0 Y^{ab} Y^{cd} + J_1 (Y^{ab} Z^{cd} + Z^{ab} Y^{cd}) 
	\nonumber
	\\
	{} + J_2 (Z^{ab} Z^{cd} - X^{ab} Y^{cd} - Y^{ab} X^{cd} )
	\nonumber
	\\
	{} -  J_3 ( X^{ab} Z^{cd} + Z^{ab} X^{cd}) + J_4 X^{ab} X^{cd} \big].
	\label{quad}
\end{align}
The complex quadrupole scalars $J_0, \ldots, J_4$ thus encode the same information as the real mass and momentum quadrupoles $(\mathscr{M}^{ab}, \mathscr{P}^{abc})$. Each complex scalar encodes two real numbers, so the number of real components needed to specify the five quadrupole scalars is $2 \cdot 5 = 10$, as expected for the number of independent components of the trace-free quadrupole moment $\mathscr{J}^{abcd}$.

If the expansions \eqref{Weyl} and \eqref{quad} are substituted into the generalized force \eqref{FgenQuad}, we find that in general,
\begin{widetext}
\begin{align}
	\gF^{\rm quad}_\xi =  \frac{4}{3} \Re \big\{ [ \Psi_0 ( J_4 X^{ab} - J_3 Z^{ab} ) \Lie_\xi Y_{ab} - J_4 \Lie_\xi \Psi_0] + \big[ \Psi_1 ( J_4 X^{ab} - 3 J_2 Y^{ab} ) \Lie_\xi Z_{ab} - 2 J_3 X^{ab} \Lie_\xi (\Psi_1 Y_{ab}) \big] 
	\nonumber
	\\
	{} + 3 \big[ \Psi_2 ( J_1 Y^{ab} - J_3 X^{ab} ) \Lie_\xi Z_{ab} - 2 J_2 \Lie_\xi \Psi_2 \big] + \big[ \Psi_3 ( 3 J_2 X^{ab} - J_0 Y^{ab} ) \Lie_\xi Z_{ab} - 2  J_1 Y^{ab} \Lie_\xi (\Psi_3 X_{ab} ) \big]
    \nonumber
    \\
    {} + \big[  \Psi_4 ( J_0 Y^{ab} + J_1 Z^{ab}) \Lie_\xi X_{ab}  - J_0 \Lie_\xi \Psi_4 \big]\big\}.
\end{align}
\end{widetext}
This is equivalent to \eqref{FgenQuad3}, which expresses the generalized force in terms of $\mathscr{M}^{ab}$ and $\mathscr{P}^{abc}$. As-is, it is considerably more complicated than that equation. However, this result simplifies when the tetrad is adapted to a spacetime's principal null directions. For example, $\Psi_0 = 0$ when $\ell^a$ is aligned with one of those directions, which eliminates one of the bracketed terms here. Further simplifications can be arranged in algebraically-special spacetimes, and their consequences have been discussed extensively in \cite{Ha.23}. The Petrov type D case is also recalled in Sec. \ref{Sec:TypeD} below.

Another point that can be noted is that the null tetrad is not unique. Various transformation may be applied to it, and those transformations induce changes in the bivector basis, the Weyl scalars, and the quadrupole scalars. These issues are discussed in \cite{Ha.23}. See also Sec. 1.8(g) of \cite{Cha}.

\subsection{Decomposing the octupole}
\label{Sec:Nulloct}

We now extend the null decomposition of the trace-free quadrupole moment $\mathscr{J}^{abcd}$ to a null decomposition of the trace-free octupole moment $\mathscr{J}^{abcde}$. By analogy with \eqref{Ji} and \eqref{quad}, which involve the five complex \textit{scalars} $J_0, \ldots, J_4$, consider the five complex \textit{vectors} 
\begin{subequations}
\label{Jioct}
\begin{align}
	J_0^a &\equiv \frac{1}{4} \mathscr{J}^{a}{}_{bcde} X^{bc} X^{de}, \quad J_1^a \equiv \frac{1}{8} \mathscr{J}^{a}{}_{bcde} X^{bc} Z^{de},
	\\
	J^a_2 &\equiv -\frac{1}{4} \mathscr{J}^{a}{}_{bcde} X^{bc} Y^{de} = \frac{1}{16} \mathscr{J}^{a}{}_{bcde} Z^{bc} Z^{de},
	\\
	J^a_3 &\equiv -\frac{1}{8} \mathscr{J}^{a}{}_{bcde} Y^{bc} Z^{de}, \quad J^a_4 \equiv \frac{1}{4} \mathscr{J}^{a}{}_{bcde} Y^{bc} Y^{de}.
\end{align}
\end{subequations}
We refer to these as the octupole vectors, and their definitions are consistent with
\begin{align}
	\mathscr{J}^{abcde} = 2 \Re \big[ J^a_0 Y^{bc} Y^{de} + J^a_1 (Y^{bc} Z^{de} + Z^{bc} Y^{de}) 
	\nonumber
	\\
	{} + J^a_2 (Z^{bc} Z^{de} - X^{bc} Y^{de} - Y^{bc} X^{de} )
	\nonumber
	\\
	{} -  J^a_3 ( X^{bc} Z^{de} + Z^{bc} X^{de}) + J_4^a X^{bc} X^{de} \big].
	\label{oct}
\end{align}
The right-hand side here is antisymmetric in the indices $bc$ and $de$, and it vanishes when antisymmetrized over the triple $cde$. It is also trace-free over all index pairs in $bcde$. However, unlike $\mathscr{J}^{abcde}$, the right-hand side is not automatically trace-free over, e.g., the indices $ab$. It might also fail to vanish when antisymmetrized over the triple $abc$. Our octupole vectors must therefore be constrained if they are to represent true octupole moments. 

In order to find those constraints, first demand that the right-hand side of \eqref{oct} be trace-free over the indices $ab$. A direct calculation  shows that this is possible only when $J_1^a$ and $J_3^a$ are related to the remaining three octupole vectors via
\begin{subequations}
\label{J1J3cov}
\begin{align}
    J^a_1 &= J_2^b X_{b}{}^{a} + J_0^b Y_{b}{}^{a}, \\
    J^a_3 &= J_4^b  X_{b}{}^{a} + J_2^b Y_{b}{}^{a} .
\end{align}
\end{subequations}
In fact, these conditions are both necessary and sufficient for the right-hand side of \eqref{oct} to be i) trace-free over all index pairs, and ii) for all symmetries in \eqref{Jtf1} to be satisfied. The three complex vectors $J_0^a$, $J_2^a$, and $J_4^a$ can thus be used to specify any trace-free octupole moment. They encode the same information as the real mass and momentum octupoles $(\mathscr{M}^{abc}, \mathscr{P}^{ab}, \mathscr{P}^{abcd})$. Given them, the remaining two octupole vectors appearing in the decomposition \eqref{oct} are fixed by \eqref{J1J3cov}. 

The three unconstrained complex 4-vectors associated with the null decomposition encode $3 \cdot 2 \cdot 4 = 24$ real independent components, which coincides with the number of independent components for $\mathscr{J}^{abcde}$ that were found using the $3+1$ decomposition in Sec. \ref{sec:31oct} above. However, the latter decomposition takes into account $t_a J^{abcde} = 0$ while former does not. This suggests that the orthogonality condition \eqref{sym3} is not actually relevant in vacuum backgrounds; it does not constrain the space of trace-free octupole moments. 

It can also be noted that the null decomposition described here applies not only to trace-free octupole moments, but also to, e.g., the gradient of the Weyl tensor. Translating our notation for that case, \eqref{oct} and \eqref{J1J3cov} imply that there exist five complex vectors $\Psi_0^a, \ldots, \Psi^a_4$ such that
\begin{align}
	\nabla^a C_{bcde} = 2 \Re \big[ \Psi^a_0 Y_{bc} Y_{de} + \Psi^a_1 (Y_{bc} Z_{de} + Z_{bc} Y_{de}) 
	\nonumber
	\\
	{} + \Psi^a_2 (Z_{bc} Z_{de} - X_{bc} Y_{de} - Y_{bc} X_{de} )
	\nonumber
	\\
	{} -  \Psi^a_3 ( X_{bc} Z_{de} + Z_{bc} X_{de}) + \Psi_4^a X_{bc} X_{de} \big],
    \label{dWeyl}
\end{align}
where
\begin{subequations}
\label{psi1psi3cov}
\begin{align}
    \Psi^a_1 &= \Psi_2^b X_{b}{}^{a} + \Psi_0^b Y_{b}{}^{a}, \\
    \Psi^a_3 &= \Psi_4^b  X_{b}{}^{a} + \Psi_2^b Y_{b}{}^{a} .
\end{align}
\end{subequations}
These last conditions may be viewed as consequences of the differential Bianchi identity $\nabla_{[a} C_{bc]de} = 0$ that holds in vacuum spacetimes. Additional information can be gained by comparing with a direct gradient of \eqref{Weyl}, which we do in Sec. \ref{Sec:TypeD} below for spacetimes with Petrov type D.

\section{Motion in Newtonian spacetimes}
\label{Sec:Newtonian}

In order to gain intuition for how an octupole moment can affect an object's motion, we now consider an extended body moving in a static, nearly-Newtonian spacetime. Working in a Newtonian framework from the start, at least quadrupolar extended-body effects have been extensively discussed in, e.g., \cite{Ha.21, Ha.23}. A somewhat different perspective is taken here: Beginning with the fully-relativistic theory, we consider a possibly-relativistic body---perhaps an electromagnetic wavepacket \cite{HaOa.22, An.26, Co.26}---that is moving in a ``Newtonian spacetime.'' By this, we mean not only that a body's center of mass might be moving rapidly with respect to a certain Newtonian frame, but also that its internal momentum densities and stresses could be significant compared with its energy density. We show that the momentum moments can then affect the motion, which is not possible in a fully-Newtonian context. They do so by inducing a hidden momentum: a misalignment between a body's velocity and its momentum. 

\subsection{Newtonian spacetimes}

If $\eta_{ab}$ denotes a flat metric and $\Phi$ a Newtonian potential, a corresponding ``Newtonian metric'' can be written as \cite{Wald}
\begin{equation}
    g_{ab} = \eta_{ab} - 2 \Phi (\eta_{ab} + 2 \nabla_a t \nabla_b t) + \mathscr{O}(\Phi^2),
    \label{gNewt}
\end{equation}
where $t$ is an inertial time coordinate for $\eta_{ab}$. We assume that the potential is static with respect to the future-directed timelike vector field 
\begin{equation}
    \tau^a \equiv -\eta^{ab} \nabla_b t,
\end{equation}
in the sense that $\Lie_\tau \Phi = 0$. Hence,
\begin{equation}
  \Lie_\tau g_{ab} = O(\Phi^2).
  \label{newtKilling}
\end{equation}
It follows that $\tau^a$ is approximately Killing in the spacetimes considered here.

The laws of motion depend on the curvature and its derivatives, and a direct calculation shows that 
\begin{align}
    R_{abcd} = 2 \big[ (\eta_{d[b} + 2 \nabla_d t \nabla_{[b} t )\nabla_{a]} \nabla_c \Phi - (\eta_{c[b} 
    \nonumber
    \\
    {} + 2 \nabla_c t \nabla_{[b} t )\nabla_{a]} \nabla_d \Phi \big] + \mathscr{O}(\Phi^2).
    \label{RNewt}
\end{align}
Enforcing the vacuum Einstein equation (with vanishing cosmological constant) $R_{ab} = 0$ through first order in $\Phi$ recovers the Newtonian field equation $\nabla^2 \Phi = 0$, where $\nabla^2$ is the ordinary Laplacian constructed from the Euclidean spatial metric
\begin{equation}
    e_{ab} \equiv \eta_{ab} + \nabla_a t \nabla_b t.
    \label{Euclid}
\end{equation}

\subsection{Conservation laws and centroids}

Recalling that the existence of any Killing field implies a conservation law for the motion of an extended body, it follows from \eqref{newtKilling} that the energy 
\begin{equation}
    \mathscr{E} \equiv \mathscr{P}_{-\tau} = -g_{ab} \left( p^a \tau^b + \tfrac{1}{2} S^{ca}  \nabla_c \tau^b \right) 
\end{equation}
is conserved through first order in $\Phi$. This can be made more explicit by noting that
\begin{equation}
    g_{bc} \nabla_a \tau^c = 2 \nabla_{[a} t \nabla_{b]} \Phi + \mathscr{O}(\Phi^2), 
    \label{dt}
\end{equation}
so
\begin{equation}
    \mathscr{E} =  [(1+2 \Phi) p^a - S^{ab} \nabla_b \Phi ]  \nabla_a t + \mathscr{O}(\Phi^2).
    \label{energyNewt}
\end{equation}
Energy conservation implies that $\gF_\tau = - \ud \mathscr{E}/\ud s = \mathscr{O}(\Phi^2)$, and because the force and the torque are already linear in $\Phi$, it follows from \eqref{genFFN} and \eqref{dt} that forces must satisfy
\begin{align}
    F^a \nabla_a t = \mathscr{O}(\Phi^2) 
    \label{Ftnewt}
\end{align}
in Newtonian spacetimes. This holds through all multipolar orders.

One more ingredient needed to understand motion in Newtonian spacetimes is the worldline $\mathcal{Z}$. This serves as an origin for the multipolar expansion, and we would like it to be interpreted as a body's centroid. It is simpler here not to enforce the centroid interpretation using the Tulczyjew-Dixon spin supplementary condition \eqref{centroid}, but instead to assume that 
\begin{equation}
    S^{ab} \nabla_b t = 0.
    \label{SSC}
\end{equation}
This implies that $\mathcal{Z}$ is to be chosen such that the mass dipole moment vanishes in the Newtonian rest frame. Similar choices have sometimes been referred to as Corinaldesi-Papapetrou spin supplementary conditions \cite{CoNa.15}. Regardless, using this condition removes the angular momentum from the conserved energy \eqref{energyNewt}:
\begin{equation}
    \mathscr{E} = (1+2\Phi) p^a \nabla_a t + \mathscr{O}(\Phi^2).
    \label{energy}
\end{equation}
Note that while $\mathscr{E}$ cannot change, the relativistic rest mass $M \equiv \sqrt{ -g_{ab} p^a p^b }$ can \cite{DiI.70, Ha.20}.

The spin supplementary condition \eqref{SSC} implicitly selects a worldline $\mathcal{Z}$. Parameterizing that worldline by $z(s)$ leaves open a choice for the worldline parameter $s$. We now identify $s$ with the Newtonian time $t$, so 
\begin{equation}
    \dot{z}^a \nabla_a t = 1.
\end{equation} 
Differentiating \eqref{SSC} while using \eqref{Sdot} and \eqref{energy} then shows that
\begin{equation}
    (1-2\Phi) \mathscr{E} \dot{z}^a = p^a - S^{ab} \nabla_b \Phi + N^{ab} \nabla_b t + \mathscr{O}(\Phi^2).
    \label{pVel}
\end{equation}
This relates the velocity of an object's centroid to its momentum, which is not necessarily parallel. Misalignments between the linear momentum and the velocity are sometimes attributed to ``hidden momenta'' \cite{Gr.al2.10, CoNa.15}. They are relativistic effects\footnote{If we had not set the speed of light to unity, both the second and the third terms on the right-hand side of \eqref{pVel} would be suppressed by powers of $c$.} that depend, in this context, on both the angular momentum $S^{ab}$ and the torque $N^{ab}$. The coupling of the angular momentum to the Newtonian acceleration $-\nabla_a \Phi$ might be described as a ``spin Hall'' effect in an optical context where a (say) electromagnetic wavepacket has nonzero angular momentum \cite{HaOa.22}. Of the six torque components encoded in $N^{ab}$, three affect the evolution of the angular momentum; the remaining three affect the hidden momentum. 

A shape-changing spacecraft would have a limited ability to control its angular momentum on short timescales. However, the torque that acts upon such a spacecraft \textit{is} directly controllable. It can be varied as rapidly as a body can vary its multipole moments. A spacecraft might therefore control its hidden momentum---and thus its velocity---by manipulating its multipole moments in order to appropriately vary $N^{ab} \nabla_b t$. We now show that it is a body's momentum moments that allow for this type of control in nearly-Newtonian spacetimes. 

\subsection{Forces and torques in static Newtonian spacetimes}

Given that there is a preferred timelike vector in Newtonian spacetimes, laws of motion are relatively simple when written in terms of that vector. We therefore employ the $3+1$ decompositions of Sec. \ref{Sec:31decomp}, choosing the $t^a$ used there to coincide with the $\tau^a$ in the Newtonian metric\footnote{Our definitions thus far have required that $g_{ab} t^a t^b = -1$ while $\eta_{ab} \tau^a \tau^b = -1$. However, this difference in normalization arises only at order $\Phi$, which is irrelevant for the force and torque expansions provided here. Similarly, the Euclidean metric $e_{ab}$ defined by \eqref{Euclid} coincides with the spatial projection $h_{ab}$ defined by \eqref{hDef} up to terms of order $\Phi$.}. The quadrupolar generalized force \eqref{FgenQuad3} then reduces to 
\begin{equation}
    \gF_\xi^{\rm quad} = - \frac{1}{2} \mathscr{M}^{ab} \Lie_\xi \nabla_a \nabla_b \Phi + \frac{8}{3} \mathscr{P}^{ab}{}_{c} (\nabla_a \nabla_b \Phi) \Lie_\xi t^c
    \label{FquadNewt}
\end{equation}
through first order in $\Phi$. The first term coincides with the ordinary Newtonian expression for the quadrupolar generalized force \cite{Ha.23}. Although the second term is negligible for a Newtonian body whose momentum density is much smaller than its energy density, it can affect more relativistic objects. 

Noting that $\Lie_\xi t^c = -t^d \nabla_{d} \xi^c + \mathscr{O}(\Phi)$ in this context, momentum quadrupoles cannot affect either the force or the space-space components of the torque. They can affect only the time-space components of the torque that contribute to the hidden momentum via \eqref{pVel}. Control of the current quadrupole therefore allows for control of an object's hidden momentum, and hence its velocity. 

All components of the quadrupolar force and torque can be extracted either by comparing \eqref{FquadNewt} with \eqref{genFFN}, or by applying \eqref{FNquad}. In either case, 
\begin{subequations}
\label{FNquadNewt}
\begin{align}
    F_a^{\rm quad} &= - \frac{1}{2} \mathscr{M}^{bc} \nabla_a \nabla_b \nabla_c \Phi,
\\
    N^{ab}_{\rm quad} &= \frac{2}{3} \big(8 \mathscr{P}^{cd[a} t^{b]} -3 \mathscr{M}^{c[a} h^{b]d} \big) \nabla_c \nabla_d \Phi.
\end{align}
\end{subequations}
The time-space components of the torque that contribute to the hidden momentum are thus
\begin{equation}
    N^{ab}_{\rm quad} \nabla_b t = \frac{8}{3} \mathscr{P}^{cd a} \nabla_c \nabla_d \Phi.
\end{equation}

The octupole case can be understood similarly. Inserting \eqref{RNewt} into \eqref{FgenOct2} shows that the generalized force is given by
\begin{align}
    \gF_\xi^{\rm oct} = \frac{1}{6} \big[ -\mathscr{M}^{(abc)} \Lie_\xi \nabla_a \nabla_b \nabla_c \Phi + \big( 6 \mathscr{P}^{(abc)}{}_{d} 
    \nonumber
    \\
    {} + \mathscr{P}^{(ab} h^{c)}{}_{d} \big) (\nabla_a \nabla_b \nabla_c \Phi ) \Lie_\xi t^d \big] .
    \label{FoctNewt}
\end{align}
Decomposing this into a force and a torque then shows that
\begin{subequations}
\label{FNoctNewt}
\begin{align}
    F_a^\text{oct} &=  -\frac{1}{6} \mathscr{M}^{(bcd)} \nabla_a \nabla_b \nabla_c \nabla_d \Phi ,
    \\
    N^{ab}_{\rm oct} &=  \frac{1}{6} \big[ 3 \big( h^{ae} \mathscr{M}^{(bcd)} -h^{be} \mathscr{M}^{(acd)} \big) 
    \nonumber
    \\
    & {}+ 2 \big(6 \mathscr{P}^{(cde) [a}  +  \mathscr{P}^{(cd} h^{e)[a}  \big) t^{b]} \big] \nabla_c \nabla_d \nabla_e \Phi.
\end{align}
\end{subequations}
Although $\mathscr{M}^{abc}$ is not necessarily symmetric in all three indices, only its fully-symmetric component appears here. The Newtonian octupole moment is also fully symmetric, and all terms  involving the relativistic mass moment exactly match standard Newtonian expressions. However, a body's momentum moments can also contribute. Those moments again appear only in the time-space components of the torque, via
\begin{align}
    N^{ab}_{\rm oct} \nabla_b t = \frac{1}{6} \big(6 \mathscr{P}^{(cde) a}  +  \mathscr{P}^{(cd} h^{e)a}  \big) \nabla_c \nabla_d \nabla_e \Phi.
\end{align}
Velocities can thus be controlled by momentum octupoles in the combination $6 \mathscr{P}^{(cde)a} + \mathscr{P}^{(cd} h^{e)a}$.

It has been noted before that a body capable of controlling its shape could control its hidden momentum in certain contexts. For example, the hidden momentum can be varied arbitrarily in both the Friedmann-Lema\^{\i}tre-Robinson-Walker \cite{Ha.07} and the Kasner \cite{Ha.23} spacetimes, despite the conservation laws associated with the spatial homogeneity of both spacetimes.

\subsection{Forces and torques around a spherical mass}
\label{Sec:newtPointMass}

As a simple illustration of our force and torque expressions in Newtonian spacetimes, we now consider their form for an extended body moving in the field of a large, spherically-symmetric object with mass $m$. Introducing a radial coordinate $r$ for the background Euclidean metric, the potential in this context is
\begin{equation}
    \Phi = - \frac{m}{r}.
\end{equation}
Spherical symmetry implies that in addition to the conserved energy $\mathscr{E}$, there are also three conserved angular momenta. Their presence implies that non-radial forces can exist only at the expense of torques; see, e.g., \cite{Ha.20, Ha.21}.

We now determine all possible forces and torques in this potential through octupolar order. First introducing the radial covector $\hat{r}_a \equiv \nabla_a r$, which has unit magnitude with respect to $\eta_{ab}$, the second and the third derivatives of $\Phi$ are given by
\begin{subequations}
    \begin{align}
        \nabla_a \nabla_b \Phi &= \frac{ m }{ r^3 } ( h_{ab} - 3 \hat{r}_a \hat{r}_b) , \\
        \nabla_a \nabla_b \nabla_c \Phi &= \frac{3m}{r^4} ( 5 \hat{r}_a \hat{r}_b \hat{r}_c - 3 \hat{r}_{(a} h_{bc)} ).
    \end{align}
\end{subequations}
Using either \eqref{FquadNewt} and \eqref{FoctNewt} or \eqref{FNquadNewt} and \eqref{FNoctNewt}, it follows that the force is given by
\begin{align}
    F_a = \frac{3 m}{ 2r^4 } ( 2 h_{ab} - 5 \hat{r}_a \hat{r}_b ) \mathscr{M}^{bc} \hat{r}_c - \frac{5 m }{ 2 r^5} \big( 3 h_{ab} 
    \nonumber
    \\
    {} - 7 \hat{r}_a \hat{r}_b \big) \mathscr{M}^{(bcd)} \hat{r}_c \hat{r}_d  + \ldots
\end{align}
through octupolar order. Similarly, the torque is equal to 
\begin{align}
    N^{ab} = \frac{6m}{r^3} \hat{r}_c \mathscr{M}^{c[a} \hat{r}^{b]} +  \frac{5 m }{ r^4 } \hat{r}^{[a} (  \mathscr{M}^{b]cd} + 2 \mathscr{M}^{|cd|b]} ) \hat{r}_c \hat{r}_d 
    \nonumber
    \\
    {} - \frac{16m}{r^3} \hat{r}_c \hat{r}_d  \mathscr{P}^{cd[a} t^{b]} + \frac{m}{r^4} [ 5 \hat{r}_c \hat{r}_d ( 6 \hat{r}_e \mathscr{P}^{cde[a}
    \nonumber
    \\
    {} + \mathscr{P}^{cd} \hat{r}^{[a} )- 2 \hat{r}_c \mathscr{P}^{c[a} \big] t^{b]} + \ldots
\end{align}
Everything here involving the mass moments $\mathscr{M}^{ab}$ and $\mathscr{M}^{abc}$ is essentially Newtonian. Some Newtonian extended-body effects have been investigated in \cite{Ha.21}, at least at quadrupolar order and in cases with vanishing torque. Even with those restrictions, it was found to be possible for a suitably-engineered object to use changes in $\mathscr{M}^{ab}$ to modify the eccentricity and the energy of its orbit, and to rotate its line of apsides. Similar results are also known for torque-free extended bodies in the Schwarzschild spacetime \cite{Ha.20}.

What appears to be new here are the contributions of the momentum moments $\mathscr{P}^{ab}$, $\mathscr{P}^{abc}$, and $\mathscr{P}^{abcd}$. Given the momentum-velocity relation \eqref{pVel}, these moments affect a body's hidden momentum  via
\begin{align}
    N^{ab} \nabla_b t = - \frac{ 8m }{r^3} \hat{r}_c \hat{r}_d  \mathscr{P}^{cda} + \frac{ m }{ 2 r^4 } [ 5 \hat{r}_c \hat{r}_d ( 6 \hat{r}_e \mathscr{P}^{cdea}
    \nonumber
    \\
    {} + \mathscr{P}^{cd} \hat{r}^{a} ) - 2 \hat{r}_c \mathscr{P}^{ac} ] + \ldots.
\end{align}
Both the quadrupole and the octupole moments can be used to produce arbitrary non-radial hidden momenta.
Noting that
\begin{align}
    \hat{r}_a (N^{ab} \nabla_b t) = \frac{ 3 m }{ 2 r^4 }  \hat{r}_a \hat{r}_b \mathscr{P}^{ab} + \ldots ,
\end{align}
it is also possible to control the radial component of the hidden momentum. However, this requires control of the octupole moment $\mathscr{P}^{ab}$; the radial hidden momentum cannot be controlled with quadrupole moments alone. This is our first example of an octupolar effect that does not arise at quadrupolar order. It is shown in Sec. \ref{Sec:TypeD} below that analogous differences are generically present in type-D spacetimes, even in fully-relativistic settings.

One more comment that can be made here is that hidden momentum effects---which contribute to an object's translational motion---fall off at large distances like $1/r^3$ for non-radial contributions. By contrast, ordinary extended-body forces fall off more rapidly, like $1/r^4$. Even though the hidden momentum is an intrinsically-relativistic effect, and therefore small in most systems, this suggests that there may be regimes in which hidden momentum effects dominate over Newtonian extended-body effects even in very weak gravitational fields.

\section{Motion in type D spacetimes}
\label{Sec:TypeD}

We now consider the motion of fully-relativistic extended bodies in fully-relativistic spacetimes. In particular, our focus is on vacuum spacetimes that are of Petrov type D, meaning that their Weyl tensors admit two doubly-degenerate principal null directions \cite{SKMHH, Cha}. Important examples in this class include the Schwarzschild and the Kerr solutions, although our discussion is not restricted only to these cases. 

We begin by reviewing the geometry of type D spacetimes, and then the quadrupole forces and torques, which largely follows the discussion in \cite{Ha.20}. Finally, we derive all possible octupolar forces and torques in vacuum type D backgrounds. It is less useful in this context to decompose a body's multipole moments in terms of a timelike vector field, as in the $3+1$ decomposition of Sec. \ref{Sec:31decomp}, so we instead adopt the null decomposition of Sec. \ref{Sec:NullDecomp}. One of our principal results is that generically, there exist torques that can be produced with octupole moments but not with quadrupole moments. 

\subsection{Geometry of type D spacetimes}

A four-dimensional spacetime is said to be of Petrov type D if its Weyl tensor admits two doubly-degenerate principal null directions. Adapting the null tetrad $(\ell^a, n^a, m^a, \bar{m}^a)$ introduced in Sec. \ref{Sec:NullDecomp} such that the real vectors $\ell^a$ and $n^a$ are aligned with these directions, all but one Weyl scalar can be shown to vanish:
\begin{equation}
    \Psi_0 = \Psi_1 = \Psi_3 = \Psi_4 = 0.
\end{equation}
The remaining Weyl scalar, $\Psi_2$, is all that is needed to determine the curvature. Using \eqref{Weyl}, that curvature is given by
\begin{equation} \label{defDWeyl}
  C_{abcd}  = 2 \Re [\Psi_2 (Z_{ab} Z_{cd} - X_{ab} Y_{cd} -Y_{ab} X_{cd})],
\end{equation}
where the antisymmetric tensors $X_{ab}$, $Y_{ab}$, and $Z_{ab}$ are constructed from the tetrad using \eqref{defXYZ}. Quadrupolar forces and torques depend on Lie derivatives of this expression with respect to the generalized Killing fields.

Octupolar forces and torques instead depend on Lie derivatives of $\nabla_a C_{bcde}$. The algebraic and differential Bianchi identities imply that this gradient  must have the form \eqref{dWeyl}, which depends on the five Weyl vectors $\Psi_0^a, \ldots, \Psi^a_4$. In the type-D context where $\ell^a$ and $n^a$ have been aligned with the principal null directions, the Weyl vectors can be computed by differentiating \eqref{defDWeyl} and then applying the $J^a_{I} \to \Psi^a_{I}$ analog of \eqref{Jioct}, which shows that
\begin{equation}
    \Psi_0^a = \Psi_4^a = 0, \qquad \Psi^a_2 = \nabla^a \Psi_2.
    \label{typeD1}
\end{equation}
The remaining Weyl vectors, $\Psi_1^a$ and $\Psi_3^a$, depend on $\Psi_2^a$ via \eqref{psi1psi3cov}. All Weyl vectors are thus determined by the second. 

There are actually two ways to relate the first and the third Weyl vectors to the second. Besides the method just mentioned,  \eqref{Jioct} can also be used to extract these quantities directly from the gradient of \eqref{defDWeyl}. Using the results of both computations,
\begin{subequations}
\label{typeD2}
\begin{align}
     \Psi^a_1 &= - X^{ab} \nabla_b \Psi_2 = -\frac{3}{4} \Psi_2 (X^{bc} \nabla^a Z_{bc}) , 
    \\
    \Psi^a_3 &= - Y^{ab} \nabla_b \Psi_2 = \frac{3}{4} \Psi_2 (Y^{bc} \nabla^a Z_{bc}) .
\end{align}
\end{subequations}
The latter equalities in both lines here may be viewed as allowing $\nabla_a \Psi_2$ to be written in terms of $\Psi_2$ itself. This can be made more explicit by first noting that it follows from the first identity in \eqref{bivectorProds} that $g^{ab} = - 2 X^{(a}{}_{c} Y^{b)c}$. Contracting both sides of that equation with $\nabla_b \Psi_2$ while using \eqref{typeD2} shows that
\begin{align}
    \nabla_a \Psi_2 = \frac{3}{4} \Psi_2 ( Y_{a}{}^{b} X^{cd} - X_{a}{}^{b} Y^{cd} ) \nabla_b Z_{cd}.
\end{align}
Although this is not used below, we include it for completeness.

\subsection{Quadrupolar forces and torques}

At quadrupolar order, forces and torques in type D spacetimes have been extensively discussed in \cite{Ha.20, Ha.23}. It was shown there that the generalized force is given by
\begin{align} \label{Fquad}
    \gF^{\rm quad}_\xi = -4 \Re \big[ 2 J_2 \Lie_{\xi} \Psi_2 + \Psi_2 (J_3 X^{ab} - J_1 Y^{ab}) 
    \nonumber
    \\
    {} \times \Lie_\xi Z_{ab} \big] ,
\end{align}
which encodes both the ordinary quadrupolar force 
\begin{align}
	F_a^{\rm quad} = -4  \Re \big[ 2 J_2 \nabla_a \Psi_2 + \Psi_2 (J_3 X^{bc} - J_1 Y^{bc} ) 
    \nonumber
    \\
    {} \times \nabla_a Z_{bc}],
	 \label{quadF}
\end{align}
as well as the quadrupolar torque
\begin{equation}
	N_{ab}^{\rm quad} = 16  \Re \left[ \Psi_2 ( J_1 Y_{ab} + J_3 X_{ab})\right] .
	\label{quadN}
\end{equation}
These expressions can be combined to write the force as an affine function of the torque:
\begin{align}
    F_a^{\rm quad} = \frac{1}{4} \Re \big[ (X^{bc} Y^{df} - Y^{bc} X^{df} ) \nabla_a Z_{bc} \big] N_{df}^{\rm quad}
    \nonumber
    \\
    {} - 8 \Re \big[ J_2 \nabla_a \Psi_2].
    \label{FaffineNquad}
\end{align}
It follows that i) the quadrupole scalars $J_0$ and $J_4$ cannot affect motion in type D spacetimes, ii) $J_2$ cannot affect the torque, iii) the torque can vary throughout the four-dimensional space constrained by 
\begin{equation}
    Z_{ab} N^{ab}_{\rm quad} = 0,
    \label{torqueConstr}
\end{equation}
and iv) the quadrupolar force can be varied independently of the torque only throughout the space spanned by the real and the imaginary components of $\nabla_a \Psi_2$. This last result implies that, depending on which type D spacetime is considered, the space of forces that are possible at fixed torque can be zero, one, or two-dimensional. We recall these results mainly for comparison with the octupolar case considered below.

 Before discussing octupolar effects, however, we first apply the identities in \eqref{typeD2} to find alternative expressions for the quadrupolar force. Employing those identities in \eqref{quadF} shows that
\begin{align}
    F_a^{\rm quad} = - \frac{8}{3} \Re \Big\{ \big[ 3 J_2 \delta^b_a + 2 (J_1 Y_{a}{}^{b} + J_3 X_{a}{}^{b} ) \big] 
    \nonumber
    \\
    {} \times \nabla_b \Psi_2 \Big\},
\end{align}
which illustrates that the force can be viewed as a linear transformation applied to $\nabla_a \Psi_2$ and its complex conjugate. Applying a similar simplification to \eqref{FaffineNquad} shows that
\begin{align}
    F_a^{\rm quad} = \frac{1}{3} \Re \big[ (X_{a}{}^{b} Y^{cd} - Y_{a}{}^{b} X^{cd} ) \nabla_b \ln \Psi_2 \big] N_{cd}^{\rm quad}
    \nonumber
    \\
    {} - 8 \Re \big[ J_2 \nabla_a \Psi_2],
    \label{FaffineNquad2}
\end{align}
when $\Psi_2 \neq 0$. The torque-dependent components of the quadrupolar force are therefore linear combinations of $X_{a}{}^{b} \nabla_b \Psi_2$, $Y_{a}{}^{b} \nabla_b \Psi_2$, and their complex conjugates. 

\subsection{Octupolar forces and torques in type D spacetimes}

We now consider octupolar forces and torques in type D spacetimes. Substituting the octupole moment \eqref{oct} and the Weyl gradient \eqref{dWeyl} into the generalized force \eqref{FgenOct} while using \eqref{typeD1},
\begin{align}
    \gF^{\rm oct}_\xi = - \frac{ 10}{3} \Re \big[ 2 J_2^a \Lie_\xi \nabla_a \Psi_2 + (J_3^a X^{bc}  - J_1^a Y^{bc})
     \nonumber
     \\
     {} \times (\nabla_a \Psi_2 )\Lie_\xi Z_{bc} \big].
     \label{gFOct}
\end{align}
Other than the prefactor and straightforward scalar-to-vector replacements, this has exactly the same form as its quadrupolar counterpart \eqref{Fquad}. Its interpretation is different, however. This is because, although there are no constraints on the quadrupole scalars $J_1$, $J_2$, and $J_3$, the octupole vectors $J_1^a$, $J_2^a$, and $J_3^a$ cannot be varied independently; they must be related via \eqref{J1J3cov}. That constraint can be rearranged to show that 
\begin{equation}
    J_2^a = J_1^b Y_{b}{}^{a} + J_3^b X_{b}{}^{a},
    \label{J2fromJ13}
\end{equation}
so $J_2^a$ may be viewed as being determined by $J_1^a$ and $J_3^a$. From that perspective, all octupole components that can affect the force and the torque are encoded in the two complex vectors $J_1^a$ and $J_3^a$, which are not constrained. These vectors determine 16 real components of $\mathscr{J}^{abcde}$. The remaining $24-16 = 8$ components of a body's trace-free octupole are irrelevant to motion in vacuum type D spacetimes. 

These irrelevant components can be identified more precisely by again noting that all trace-free octupole moments may be parameterized by $J_0^a$, $J_2^a$, and $J_4^a$. Given \eqref{J2fromJ13}, the second octupole vector clearly contributes to motion in at least some type D spacetimes. However, use of \eqref{defXYZ} and \eqref{J1J3cov} shows that the four complex components
\begin{gather}
    J_0^a \ell_a, \quad J_0^a m_a, \quad J_4^a n_a, \quad J_4^a \bar{m}_a
\end{gather}
of $J_0^a$ and $J_4^a$ do not contribute either to $J_1^a$ or to $J_3^a$. These are the irrelevant components of the trace-free octupole moment. 

Regardless, the octupolar force and the octupolar torque can now be extracted from the generalized force \eqref{gFOct} using \eqref{genFFN}. Doing so results in
\begin{align}
    F_a^{\rm oct} = - \frac{ 10}{3} \Re \big[2 J_2^b \nabla_a \nabla_b \Psi_2 + (X^{bc} J_3^d 
     - Y^{bc} J_1^d ) 
     \nonumber
     \\
     {} \times (\nabla_d \Psi_2) \nabla_a Z_{bc} \big],
     \label{forceOct}
\end{align}
and
\begin{align}
    N^{ab}_{\rm oct} =  \frac{40}{3} \Re \big[  (J^c_1 Y^{ab} + J^c_3 X^{ab} ) \nabla_c \Psi_2 - J_2^{[a} \nabla^{b]} \Psi_2 \big].
    \label{torqueOct}
\end{align}
While this force looks very similar to its quadrupolar counterpart \eqref{quadF}, the $J_2^{[a} \nabla^{b]} \Psi_2$ term in the octupolar torque has no quadrupolar analog. That term also has an important physical consequence: Although it follows from \eqref{torqueConstr} that quadrupolar torques can vary only within a four-dimensional space, more is possible with octupole moments (except in cases where extra torques are precluded by the presence of Killing fields). Contracting \eqref{torqueOct} with $Z_{ab}$ while using \eqref{J2fromJ13} shows that
\begin{align}
    Z_{ab} N^{ab}_{\rm oct}  &= - \frac{40}{3} \Re \big[ J_2^{[a} \nabla^{b]} \Psi_2 \big] Z_{ab}, 
    \label{NZoct}
\end{align}
which does not necessarily vanish. However, \eqref{torqueConstr} shows that the same torque component always vanishes at quadrupolar order. \textit{Some motions that are possible in the octupole approximation are therefore impossible in the quadrupole approximation}. This is true in generic Kerr spacetimes (with nonzero angular momentum), for example, where $Z_{ab} N^{ab}_{\rm oct}$ can---for a body with appropriate octupole moments---take on any complex value. Octupole moments thus provide two real degrees of freedom in that case that do not exist at quadrupolar order. 

This is not true in Schwarzschild spacetimes, where the conservation laws associated with spherical symmetry show that $\Im (Z_{ab} N^{ab} ) = 0$ \cite{Ha.20}. This result is exact, and therefore holds through all multipolar orders. The real component of $Z_{ab} N^{ab}$ can nevertheless be nonzero at octupolar---but not quadrupolar---order. For an object moving in a Schwarzschild spacetime, octupole moments thus provide one real degree of freedom that does not exist at the quadrupole level. This can also be seen, in a weak-field limit, from the results of Sec. \ref{Sec:newtPointMass}, where an extra torque component available at octupole order allows for control over the radial component of a body's hidden momentum in a spherically-symmetric spacetime.

As an even more special example, consider the Nariai or anti-Nariai spacetimes, which have line elements
\begin{equation}
    \ud s^2 = - 2 \ud u \ud v + \Lambda v^2 \ud v^2 + \frac{ \ud x^2 + \ud y^2}{[1 + \frac{1}{4} \Lambda (x^2+y^2)]^2}.
\end{equation}
These are type D spacetimes that satisfy the vacuum Einstein equation with cosmological constant $\Lambda$. They are precisely the type D spacetimes in which $\Psi_2$ is  constant \cite{Bar.15}, so it follows from \eqref{gFOct}  that $\gF_\xi^{\rm oct} = 0$; there are no octupolar effects at all in Nariai or anti-Nariai spacetimes. By contrast, it has been noted that at quadrupolar order, the force vanishes and the torque is arbitrary except for the constraint \eqref{torqueConstr} \cite{Ha.23}. In fact, the six Killing fields admitted by these spacetimes can be used to show that \textit{through all orders}, $F_a = 0$ and $N^{ab} Z_{ab} = 0$. In Nariai and anti-Nariai spacetimes, nothing appears at octupolar or higher orders that is not already present at quadrupolar order. This is an exceptional case where octupole moments do not allow more than quadrupole moments.

It was shown in \cite{Ha.23} that the quadrupolar torque constraint \eqref{torqueConstr} could be viewed as a consequence of a ``local symmetry,'' which refers to a vector field $\psi^a$ that satisfies $\Lie_\psi C_{abcd} = 0$ at least at one point on the reference worldline $\mathcal{Z}$. There could be different vector fields satisfying this relation at different points on $\mathcal{Z}$, so there is no need for $\psi^a$ to be an ordinary curvature collineation. In fact, it was shown the conformal Killing-Yano tensors admitted by type D spacetimes are sufficient to guarantee the existence of local symmetries. The analogous condition at octupolar order would be $\Lie_\psi \nabla_a C_{bcde} = 0$, which is much more difficult to satisfy in the absence of a genuine Killing field. Indeed, we have shown here that the non-Killing constraints present at quadrupolar order are generically removed at octupolar order, leaving only the restrictions imposed by Killing fields.

Another question that can be addressed here is to what extent octupolar forces can be controlled independently of torques. Unlike in the quadrupolar case, this is not answered exhaustively. Instead, we find a class of octupole moments that are sufficient to guarantee a vanishing torque while potentially allowing for a nonzero force. To begin, suppose that
\begin{equation}
    J_2^a = \kappa \nabla^a \Psi_2,
\end{equation}
where $\kappa$ is any (possibly complex and possibly time dependent) scalar. It then follows from \eqref{NZoct} that $Z_{ab} N^{ab}_{\rm oct} = 0$. Use of \eqref{J1J3cov} and \eqref{torqueOct} shows that all torque components vanish if, in addition, $J_0^a$ and $J_4^a$ are chosen such that
\begin{equation}
    J_0^a Y_{a}{}^{b} \nabla_b \Psi_2 = J_4^a X_{a}{}^{b} \nabla_b \Psi_2 = 0.
\end{equation}
This is always possible. Making these choices, substitution into \eqref{forceOct} shows that
\begin{equation}
    F_a^{\rm oct} = - \frac{10}{3} \Re \big[ \kappa \nabla_a (| \nabla \Psi_2|^2) \big],
\end{equation}
where $|\nabla \Psi_2|^2 \equiv \nabla^b \Psi_2 \nabla_b \Psi_2$. The force can therefore be varied independently of the torque (at least) throughout the space spanned by the real and the imaginary components of $\nabla_a (|\nabla \Psi_2|^2)$. Comparison with \eqref{FaffineNquad} shows that this is qualitatively similar to the quadrupole case, where the force at fixed torque can be any linear combination of the real and the imaginary components of $\nabla_a \Psi_2$. In both cases, torque-free forces can be written in terms of gradients of appropriate scalars, either $\Psi_2$ or $|\nabla \Psi_2|^2$.

\section{Discussion}

This paper has explored various aspects of extended-body motion in general relativity, emphasizing octupolar contributions. Generalizing what had already been known in the quadrupole case, traces of Dixon's octupole moment were shown not to affect either forces or torques in spacetimes that satisfy the vacuum Einstein equation. This reduces the number of relevant octupole components from 40 to 24. If the background is not only vacuum, but also of Petrov type D, at least 8 additional octupole components were shown to decouple from the laws of motion. Aligning a null tetrad with the principal null directions of a type D background, the 16 real octupole components that remain can be encoded in the two complex ``octupole vectors'' $J_1^a$ and $J_3^a$ defined by \eqref{Jioct}. More generally, at most three unconstrained complex vectors (encoding 24 real components) are needed to describe all octupole components that appear in the laws of motion in generic vacuum backgrounds. This contrasts with the five complex scalars (encoding 10 real components) needed to describe generic quadrupole moments in vacuum backgrounds. 

Both of these complex decompositions depend on a choice of null tetrad. We have also performed decompositions with respect to a real timelike vector, which resulted in trace-free quadrupole and octupole moments being written in terms of trace-free mass and momentum components; see \eqref{quadTF31} and \eqref{JsplitTF}. Although the stress components $S^{abcd}$ and $S^{abcde}$ of the full (not necessarily trace-free) multipole moments can contribute to the trace-free moments, their influence is effectively absorbed into the trace-free mass moments $\mathscr{M}^{ab}$ and $\mathscr{M}^{abc}$. In the quadrupole case, \eqref{MtfQuad} shows that the contributions of the full mass and stress moments to the trace-free mass quadrupole $\mathscr{M}^{ab}$ are completely degenerate. The octupole case is more complicated, as it follows from \eqref{MtfOct} that although $\mathscr{M}^{(ab)c}$ has fully degenerate contributions from both the mass and the stress octupoles, $\mathscr{M}^{[ab]c}$ depends only on $S^{abcde}$. In both the quadrupole and the octupole cases, we find equal numbers of trace-free mass and momentum components; these have 5 components each in the quadrupole case and 12 components each in the octupole case.

This second component count differs from what might have been expected from, e.g., the multipole moments that appear in post-Newtonian theory. In that context, one also has both mass- and current-type source moments. However, both of those sets are described by fully symmetric and trace-free tensors \cite{PoWi, DSX.91, RaFl.05, Bl.14}, unlike our octupolar mass and current moments $\mathscr{M}^{abc}$ and $\Pi^{abc}$. A trace-free mass or current octupole in post-Newtonian theory would thus contain 7 independent components, not 12. It is unclear whether this difference is structural or merely terminological. The comparison may not, e.g., be between like objects.
We nevertheless leave its explanation for future work.

One of our main physical results is that a relativistic object moving in a Newtonian spacetime can manipulate its hidden momentum---and thus its velocity---by controlling its momentum moments. This provides a potential class of examples for the ``swimming in spacetime'' effect discussed by Wisdom \cite{Wi.03}, but corrected by Andrade e Silva and collaborators \cite{Silva.al.16, Silva.22}. In that context, a body's (say) displacement over a particular cyclic deformation would depend only on the sequence of shapes---or multipole moments---in that cycle, and not on their rates of change. The effective irrelevance of time here would be analogous to the swimming of microorganisms at low Reynolds numbers, where inertia is negligible \cite{Pu.77, ShWi.89, Lauga}. Although some types of ``rocket-free'' maneuvering are possible in Newtonian gravity, this swimming effect is not, and it would be interesting to understand it better. Nevertheless, we have only found which forces and torques are possible. It remains to solve the equations of motion in different circumstances and to understand their implications. 

Another one of our main physical results is that some torques that are possible with octupole moments are not possible with quadrupole moments, except where forbidden by Killing fields. Octupolar couplings can thus result in qualitatively-different motions from quadrupolar couplings. In vacuum spacetimes that are of Petrov type D, there are generically two extra torque components that can arise at octupolar order. However, this reduces to only a single component in the case of the Schwarzschild spacetime. In a Newtonian limit, this extra degree of freedom allows for control over the radial component of a body's hidden momentum. By contrast, quadrupole moments can contribute only to non-radial components of the hidden momentum. It would be interesting to understand precisely what this extra degree of freedom could be used to accomplish over long timescales. 

Another way this work could be extended would be to work to one higher order, asking how hexadecapole moments affect motion. This would not merely be a technicality, as there may be qualitative differences between octupolar and hexadecapolar effects. For example, we suggest in Sec. \ref{Sec:NoTrace} that the irrelevance of quadrupolar and octupolar traces does not necessarily extend to higher orders: Traces of a body's hexadecapole moment \textit{can} affect its motion, a conclusion that also appears to be supported by the results recently reported in \cite{AmMoVi.26}. Another interesting property of hexadecapolar generalized forces is that they can partially mix with quadrupolar generalized forces, in the sense that both can involve terms proportional to $\Lie_\xi R_{abcd}$. This and other peculiarities are due to the fact that the tensor extensions of the metric are not linear in the Riemann tensor beyond third order; see, e.g., \eqref{g4extend}.

\appendix

\section{Notation}
\label{App:Notation}

We assume throughout this paper that the spacetime has signature $(-,+,+,+)$ and that the sign of the Riemann tensor is chosen such that $2\nabla_{[a} \nabla_{b]} \omega_c = R_{abc}{}^{d} \omega_d$ for any $\omega_c$. Overbars are used to denote complex conjugates. Symbols commonly used in the text are summarized in Table \ref{Table:Notation}, except for those used to denote multipole moments, which are summarized in Table \ref{Table:Moments} above. 

In cases where a multipole moment is denoted using both ordinary and calligraphic fonts (such at $J^{abcd}$ and $\mathscr{J}^{abcd}$), the latter denotes the trace-free component of the former. Quadrupole and octupole moments are frequently denoted by the same base symbol and are distinguished only by their number of indices. An octupole typically carries one more index than its quadrupolar counterpart, such as with the mass quadrupole $M^{ab}$ and the mass octupole $M^{abc}$. However, an exception arises for the trace-free momentum moments: While the quadrupole $\mathscr{P}^{abc}$ has three indices, the octupole is described using a combination of the two tensors $\mathscr{P}^{ab}$ and $\mathscr{P}^{abcd}$. 


\begin{table*}
  \centering
  \setlength{\tabcolsep}{9pt}

  \begin{tabular}{p{12em} p{0.4\textwidth} p{12em}}
    \hline\hline
    Symbol & Description & Reference \\[.3ex]
    \hline
    $\mathcal{Z}$   &   Reference worldline    & Fig. \ref{Fig:worldtube}
    \\
    $z(s)$	&	Point on $\mathcal{Z}$ & 	Fig. \ref{Fig:worldtube}
    \\
    $\Sigma(s)$  & Leaf of worldtube foliation   & Fig. \ref{Fig:worldtube} 
    \\
    $t^a (s)$   & Vector orthogonal to $\Sigma(s)$ & Fig. \ref{Fig:worldtube}
    \\[.2ex]
    \hline
    $g_{ab} (x)$		&	Spacetime metric	&	-
    \\
    $h_{ab} (x)$		&	Spatial projection	&	\eqref{hDef}
    \\
    $\Phi (x)$      &   Newtonian potential & \eqref{gNewt}
    \\
    $\Lambda$	&	Cosmological constant	& 	\eqref{Einstein}
    \\
    $R_{abcd}(x)$, $C_{abcd}(x) $ &	 Riemann and Weyl tensors & \eqref{Weyl}
    \\
    $\xi^a(x) $		&	Generalized or ordinary Killing vector	& \eqref{genKilling}
    \\
    $\mathscr{P}_\xi(s)$	& Generalized momentum associated with $\xi^a$	& \eqref{Pdef}, \eqref{genPpS}
    \\
    $\mathscr{F}_\xi(s)$	& Generalized force associated with $\xi^a$	& \eqref{genF}, \eqref{genFFN}
    \\
    $p_a(s)$, $S^{ab}(s)$	&	Linear and angular momenta & \eqref{Dixon}, \eqref{genPpS}
    \\
    $F_a(s)$, $N^{ab}(s)$	&	Force and torque  & \eqref{Dixon}, \eqref{genFFN}
    \\[.2ex]
    \hline
    $\ell^a(x)$, $n^a(x)$, $m^a(x)$, $\bar{m}^a(x)$ & Null tetrad &	\eqref{orthonlm} 
    \\
    $X^{ab}(x)$, $Y^{ab}(x)$, $Z^{ab}(x)$ & Bivector basis elements & \eqref{defXYZ}
    \\
    $\Psi_0(x), \ldots, \Psi_4(x)$ & Weyl scalars used to decompose $C_{abcd}$ & \eqref{PsiDef}, \eqref{Weyl}
    \\
    $\Psi_0^a(x), \ldots, \Psi_4^a(x)$ & Vectors used to decompose $\nabla_a C_{bcde}$ & \eqref{dWeyl}, \eqref{psi1psi3cov}
    \\[.2ex]
    \hline\hline
  \end{tabular}
  
  \caption{Table of symbols. This excludes multipole moments, which are listed in Table \ref{Table:Moments}. Symbols listed as depending on $x$ are defined at arbitrary points in a neighborhood of a body's worldtube. Those listed as depending on $s$ are instead defined only along the reference worldline $\mathcal{Z}$ [except $\Sigma(s)$, which is a hypersurface that passes through $z(s)$].}
  \label{Table:Notation}
\end{table*}

\section{Trace-free projections for rank-5 tensors}
\label{App:TF}

Motivated by the problem of finding the trace-free component $\mathscr{J}^{abcde}$ of an octupole moment $J^{abcde}$, this appendix derives the trace-free projection $\mathscr{K}^{abcde}$ of any rank-5 tensor $K^{abcde}$  that satisfies
\begin{subequations}
\label{Ksyms}
\begin{gather}
    K^{abcde} = K^{a[bc]de} = K^{abc[de]},
    \\
    K^{[abc]de} = K^{ab[cde]} = 0.
    \label{Ksyms2}
\end{gather}
\end{subequations}
In particular, we seek a linear operator from the space of all such tensors into itself such that the result is trace-free with respect to the metric $g_{ab}$. Although we use the result only in three and four dimensions in the body of the paper, we now consider trace-free projections in generic dimension $d$. 

To begin, introduce two as-yet unknown tensors, $k_{abc} = k_{a(bc)}$ and $l_{abc} = l_{a[bc]}$, and consider the ansatz
\begin{align}
	\mathscr{K}^{abcde} = K^{abcde} - g^{a[b} l^{c]de} - g^{a[d} l^{e]bc} 
	\nonumber
	\\
	{} +  k^{ae[b} g^{c]d} - k^{a d[b} g^{c]e} .
	\label{Kdef}
\end{align}
This automatically satisfies $\mathscr{K}^{abcde} = \mathscr{K}^{a[bc]de} = \mathscr{K}^{abc[de]} = \mathscr{K}^{adebc}$. We must choose $k_{abc}$ and $l_{abc}$ in order to guarantee that this expression is trace-free, and that it satisfies Bianchi-like identities $\mathscr{K}_{[abc]de} = \mathscr{K}_{ab[cde]} = 0$.

Antisymmetrizing and then taking traces of the right-hand side shows that both Bianchi-like identities are satisfied when
\begin{equation}
	l^{abc} = 2 k^{[bc]a}.
	\label{EfromL}
\end{equation}
Substituting this into \eqref{Kdef}, it can now be observed that given any vector $v^a$, the transformation
\begin{equation}
    k^{abc} \mapsto k^{abc} + g^{a(b} v^{c)}
    \label{kGauge}
\end{equation}
cannot affect $\mathscr{K}^{abcde}$. We cannot therefore expect $k^{abc}$ to be determined uniquely. Nevertheless, an appropriate constraint may be found by demanding that all traces of $\mathscr{K}^{abcde}$ vanish.

There are essentially two types of trace that must be removed: $K^{abdc}{}_{d}$ and $K^{abcd}{}_{a}$. Both of these traces are, however, related: Use of \eqref{Ksyms} shows that
\begin{equation}
	K^{abcd}{}_{a} = -2 K^{[bc]ed}{}_{e},
\end{equation}
and
\begin{equation}
	K^{abc}{}_{bc} = 2 K^{bac}{}_{bc}.
\end{equation}
There is therefore only one trace of \eqref{Kdef} we need to remove, and doing so results in
\begin{align}
    K^{abdc}{}_{d} 
    = \frac{d}{2} k^{abc} - k^{(bc)a} + g^{ab} k^{[cd]}{}_{d} + g^{ac} k^{[bd]}{}_{d} 
    \nonumber
    \\
    {} + \frac{1}{2} g^{bc} k^{ad}{}_{d}.
\end{align}
Recalling \eqref{kGauge}, it is sufficient to introduce an ansatz in which $k^{abc}$ is a linear combination of $K^{abdc}{}_{d}$, $K^{(bc)da}{}_{d}$, and $K^{ade}{}_{de} g^{bc}$. Doing so shows that if $d \neq 2$,
\begin{align}
    k^{abc} = \frac{ 2 [ (d-1) K^{abdc}{}_{d} + 2 K^{(bc)da}{}_{d}] - K^{ade}{}_{de} g^{bc} }{ (d+1)(d-2) }.
    \label{Adef}
\end{align}
Together with \eqref{Kdef} and \eqref{EfromL}, this provides a general prescription for removing traces from rank-5 tensors that satisfy \eqref{Ksyms}.

\section{Spin-induced multipole moments}
\label{App:inducedMoments}

One common model for astrophysically-relevant multipole moments assumes that they depend only on a body's angular momentum. This is motivated by the fact that, at least in Newtonian physics and in the absence of rotation, a stationary and isolated self-gravitating fluid is expected to be spherically symmetric. The trace-free multipole moments of its density would therefore vanish. However, a rigidly-rotating fluid in equilibrium is expected to be oblate and axisymmetric (at least for small angular velocities \cite{Ch.69}), and to have nontrivial moments that depend on its angular momentum. Similar results are expected to hold in general relativity, and this appendix discusses how the multipolar decompositions of Secs. \ref{Sec:31decomp} and \ref{Sec:NullDecomp} behave for spin-induced relativistic quadrupole and octupole moments that have been discussed in the literature.

\subsection{Spin vectors and spin supplementary conditions}

Even in a Newtonian context, the trace-free moments of a spherically-symmetric body vanish only when the origin for those moments is chosen to coincide with a body's center of mass. The choice of origin is also important in relativistic settings, although there are then many possible centroid (or spin supplementary) conditions \cite{CoNa.15}. We choose here to select the worldline $\mathcal{Z}$ by applying the Tulczyjew-Dixon condition \eqref{centroid}, and to construct the foliation by setting
\begin{equation}
    t^a = p^a/M,
\end{equation}
where $M \equiv \sqrt{- p_a p^a}$ again denotes the rest mass. We assume in this appendix that $M > 0$. 

The spin supplementary condition implies that the angular momentum tensor $S^{ab}$ must be orthogonal to $p_a$, and thus has at most three independent components. Those components can be  encoded in the spin vector
\begin{equation}
    S^a \equiv - \frac{1}{2} \epsilon^{abcd} t_b S_{cd},
\end{equation}
which is orthogonal to $t_a$ (or equivalently $p_a$), and is therefore spacelike. The angular momentum tensor can be written in terms of the spin vector via
\begin{equation}
    S^{ab} = \epsilon^{abcd} t_c S_d.
    \label{Stensvect}
\end{equation}
The spin-induced multipole moments considered below depend on the squared spin tensor
\begin{equation}
    \Theta^{ab} = \Theta^{(ab)} \equiv S^{ac} S_{c}{}^{b},
    \label{thetaDef}
\end{equation}
which can alternatively be written in terms of the spin vector:
\begin{equation}
    \Theta^{ab} = S^a S^b - S^c S_c h^{ab} .
    \label{thetaExpand}
\end{equation}

Formulae below that involve $S^{ab}$ and $\Theta^{ab}$ can be simplified using the fact that\footnote{This holds not only for angular momentum tensors, but for all antisymmetric rank-2 tensors in four dimensions.}
\begin{equation} \label{IdOmega}
  S^{a b} S^{c d} + 2 S^{c [a} S^{b] d} = - 
  \frac{1}{4} (S^{e f} S^{\star}_{e f}) \epsilon^{a b c
  d},
\end{equation}
where $S_{ab}^\star \equiv \frac{1}{2} \epsilon_{abcd} S^{cd}$ denotes the Hodge dual of $S_{ab}$. This identity comes from noting that the left-hand side is totally anti-symmetric and must therefore be proportional to $\epsilon^{a b c d}$. Contraction with $\epsilon_{a b c d}$ then determines the proportionality factor. Using the result while noticing that $S^{a b} S^{\star}_{a b} = 0$ due to the spin supplementary condition \eqref{centroid}, it follows that
\begin{equation}
  S^{a b} S^{c d} = - 2 S^{c [a} S^{b] d} . \label{IdSpin}
\end{equation}
This was also noted in \cite{Di.79}, but using a different argument. Regardless, contracting \eqref{IdSpin} with $S^e{}_{c}$ while recalling the definition \eqref{thetaDef} shows that
\begin{equation}
    S^{e [a} \Theta^{b] c} =  - \frac{1}{2} S^{a b}\Theta^{e c} . \label{IdSpinTheta}
\end{equation}

\subsection{Spin-induced multipole moments}

The spin-induced quadrupole moment is expected to be given by \cite{Ma.15, Stei.11}
\begin{equation}\label{quadSpin}
    J^{abcd} = \frac{ 3 \kappa_{\rm quad} }{ M } t^{[a} \Theta^{b][c} t^{d]},
\end{equation}
which depends on the dimensionless deformability parameter $\kappa_{\rm quad}$ and on the squared-spin tensor $\Theta^{ab}$ defined by \eqref{thetaDef}. While the spin-induced quadrupole moment is expected to be quadratic in the spin, the spin-induced octupole moment is expected to be cubic. More specifically, it has been suggested to be given by \cite{Ma.15}
\begin{align} 
    J^{abcde} = \frac{\kappa_{\rm oct}}{4M^2} \big(\Theta^{a [b} S^{c] [d} t^{e]} - \Theta^{a [b} t^{c]} S^{de}    
    \nonumber 
    \\
    {} + S^{a [b} \Theta^{c] [d} t^{e]} + \Theta^{a [d} S^{e] [b} t^{c]} 
    \nonumber
     \\
     {} - \Theta^{a [d} t^{e]} S^{bc} + S^{a [d} \Theta^{e] [b} t^{c]} \big),
\end{align}
where $\kappa_{\rm oct}$ is another dimensionless deformability parameter. This satisfies the orthogonality constraint $t_a J^{abcde} = 0$. Applying the identity \eqref{IdSpinTheta} shows that it can be simplified to
\begin{equation}
    J^{abcde} = -\frac{\kappa_{\rm oct}}{2M^2} \big( S^{bc}\Theta^{a[d}t^{e]}+S^{de}\Theta^{a[b}t^{c]} \big).
    \label{octSpin}
\end{equation}
Unlike $p_a$ or $S^{ab}$, the higher-order multipole moments depend on the parameterization of the reference worldline $\mathcal{Z}$; changing $s$ rescales $\kappa_{\rm quad}$ and $\kappa_{\rm oct}$. In this case, the values $\kappa_{\rm quad} = \kappa_{\rm oct} = 1$ are expected for a black hole if the worldline is parameterized such that $p_a \dot{z}^a = - M$ \cite{Hansen:1974zz,Poisson:1997ha,Scheopner:2023rzp}. Larger values are generically expected for less compact objects and depend on the equation of state for material bodies, see \cite{FlHi.08,Rahman:2021eay,Ra.Iso.Dru.IntegO2.26,RahShaPouMat.26} and references therein. 

The physical interpretations of these expressions are somewhat unclear. Both spin-induced moments vanish with vanishing angular momentum, which is not realistic for a (say) nonspinning, spherical body. It is only the \textit{trace-free} moments $\mathscr{J}^{abcd}$ and $\mathscr{J}^{abcde}$ that might be expected to vanish for isolated and nearly stationary self-gravitating fluid bodies. However, the moments \eqref{quadSpin} and \eqref{octSpin} have nonzero traces when $S^{ab} \neq 0$. We thus expect that the expressions given here are not good models for the \textit{full} moments $J^{abcd}$ and $J^{abcde}$ of realistic bodies. What is much more likely is that their trace-free components can be used to model reasonable trace-free quadrupoles and octupoles. This distinction is, however, irrelevant when the moments are used only in vacuum backgrounds and in the laws of motion.

\subsection{Mass, momentum, and stress moments}

Regardless of physical interpretation, we now take the given spin-induced moments and perform the $3+1$ decompositions of Sec. \ref{Sec:31decomp} with respect to the timelike unit vector $t^a = p^a/M$. Applying the decomposition \eqref{quad31} to the spin-induced quadrupole \eqref{quadSpin} shows that it has only a nontrivial mass component:
\begin{equation}
    M^{ab} = -\frac{\kappa_{\rm quad}}{ M }   \Theta^{ab} , \qquad P^{abc} = S^{abcd} = 0.
\end{equation}
Applying the decomposition \eqref{Jsplit} to the spin-induced octupole \eqref{octSpin} shows that it is instead determined entirely by its momentum component:
\begin{equation}
    P^{abcd} = -\frac{\kappa_{\rm oct}}{3M^2}   \Theta^{ab}S^{cd} , \qquad M^{abc} = S^{abcde} = 0.
\end{equation}

One can also find the mass and momentum moments associated with the trace-free quadrupole and octupole moments. In the quadrupole case, \eqref{MtfQuad} and \eqref{PtfQuad} recover the expected trace-free mass moment
\begin{align}
    \mathscr{M}^{ab} = - \frac{ \kappa_{\rm quad} }{M} (\Theta^{ab})_{ \rm{TF} } = - \frac{ \kappa_{\rm quad} }{M} (S^a S^b)_{ \rm{TF} },
    \label{MtfspinInduced}
\end{align}
as well as $\mathscr{P}^{abc} = 0$. In the octupole case, applying \eqref{PtfOct}, \eqref{PtfOct2}, and \eqref{MtfOct} shows that $\mathscr{M}^{abc} = \mathscr{P}^{ab} = 0$ and 
\begin{align}
    \mathscr{P}^{abcd} = - \frac{ \kappa_{\rm oct} }{ 3  M^2 } \bigg[ \Theta^{ab} S^{cd} + \frac{1}{5} (S^{ef} S_{ef}) ( h^{ab} S^{cd} 
    \nonumber
    \\
    {} - h^{a[c} S^{d]b} - h^{b[c} S^{d]a} ) \bigg].
\end{align}

In most applications, all that is actually relevant is the generalized force. Using \eqref{FgenQuad2} and \eqref{OctForce1},
\begin{subequations}
\begin{align}
    \gF^{\rm quad}_\xi &=  \frac{ \kappa_{\rm quad} }{ 2 M } t^a S^b t^c S^d \Lie_\xi C_{abcd}, \\
    \gF^{\rm oct}_\xi &=  -\frac{ \kappa_{\rm oct} }{ 6M^2 } S^a t^b S^c t^d S^e \Lie_\xi \nabla_a C_{bcde}^\star,
\end{align}
\end{subequations}
for the spin-induced multipole moments, where $C_{bcde}^{\star} \equiv \tfrac{1}{2}\epsilon^{fg}{}_{de} C_{bc fg}$ denotes the Hodge dual of the Weyl tensor applied to its second pair of indices (which is equal to the Hodge dual applied to the first pair of indices).

\subsection{Quadrupole scalars and octupole vectors}

The null decompositions discussed in Sec. \ref{Sec:NullDecomp} can also be applied to the spin-induced multipole moments. In the quadrupole case, combining \eqref{bivectorProds} and \eqref{Ji} with the spin-induced moment \eqref{quadSpin}, the trace-free projection \eqref{quadTF} and the spin decompositions \eqref{Stensvect}-\eqref{thetaExpand}, the quadrupole scalars can be shown to be
\begin{subequations}
    \begin{align}
        J_0 &= -\frac{ 3 \kappa_{\rm quad}}{ 4 M} (x \cdot S)^2,
        \\
        J_1 &= -\frac{ 3 \kappa_{\rm quad}}{ 8 M} (x \cdot S)(z \cdot S),
        \\
        J_2 &= \frac{ \kappa_{\rm quad}}{ 8 M} \big[ 2(x \cdot S)(y \cdot S) - (z \cdot S)^2 \big],
        \\
        J_3 &= \frac{ 3 \kappa_{\rm quad}}{ 8 M} (y \cdot S) (z \cdot S), 
        \\
        J_4 &= -\frac{ 3 \kappa_{\rm quad}}{ 4 M} (y \cdot S)^2,
    \end{align}
\end{subequations}
where 
\begin{align}
    x^a \equiv X^{a}{}_{b} t^b , \qquad y^a \equiv Y^{a}{}_{b} t^b , \qquad z^a \equiv Z^{a}{}_{b} t^b
\end{align}
are complex vectors orthogonal to $t^a$. The quadrupole scalars here are all proportional to pairs of spin vector components projected into the spatial basis $(x^a, y^a, z^a)$. Some of these scalars can be written in terms of others. 

One can also compute the octupole vectors. Using \eqref{Projector4D}, \eqref{Jioct}, and \eqref{octSpin}, as well as \eqref{Stensvect}-\eqref{thetaExpand} and the fact that each of the basis bivectors is self-dual in the sense that, e.g., $X_{ab}^\star = i X_{ab}$ \cite{Ha.20}, 
\begin{subequations}
\begin{align}
    J_0^a &= \frac{ i \kappa_{\rm oct}}{ 10 M^2 }
     (S \cdot x) \big[ 5 (S \cdot x) S^a - 2 (S \cdot S) x^a \big],
    \\
    J_2^a &= \frac{ i \kappa_{\rm oct} }{ 20 M^2 } \big\{  2 \big[ (S \cdot S) - 5 (x \cdot S) (y \cdot S) \big] S^a   
    \nonumber
    \\
    & \qquad \qquad \qquad \qquad \qquad  {} - (S \cdot S) (z \cdot S) z^a \big\},
    \\
     J_4^a &= \frac{ i \kappa_{\rm oct}}{ 10 M^2 }
     (S \cdot y) \big[ 5 (S \cdot y) S^a- 2 (S \cdot S) y^a \big].
\end{align}
\end{subequations}
These vectors are all orthogonal to $t^a$. The remaining octupole vectors $J_1^a$ and $J_3^a$ can be obtained from them using \eqref{J1J3cov}.

\bibliography{ListeRef.bib}

\end{document}